**Article type**

Review

**Paper title**

RNA-seq data science: from raw data to effective interpretation

**Authors and affiliations**

† These authors contributed equally to the work.

Dhrithi Deshpande†

Department of Pharmacology and Pharmaceutical Sciences, School of Pharmacy, University of Southern California, 1985 Zonal Avenue, Room 713. Los Angeles, CA 90089-9121, USA

ddeshpan@usc.edu

Karishma Chhugani†

Department of Pharmacology and Pharmaceutical Sciences, School of Pharmacy, University of Southern California, 1985 Zonal Avenue, Room 713. Los Angeles, CA 90089-9121, USA

chhugani@usc.edu

Yutong Chang

Department of Pharmacology and Pharmaceutical Sciences, School of Pharmacy, University of Southern California, 1985 Zonal Avenue, Room 713. Los Angeles, CA 90089-9121, USA

yutongch@usc.edu

Aaron Karlsberg

Department of Clinical Pharmacy, School of Pharmacy, University of Southern California, 1540 Alcazar Street, Los Angeles, CA 90033, US

akarlsbe@usc.edu

Caitlin Loeffler
Department of Computer Science, University of California, Los Angeles, 404 Westwood Plaza, Los Angeles, CA 90095, USA
cloeffler@ucla.edu

Jinyang Zhang
Beijing Institutes of Life Science, Chinese Academy of Sciences, Beijing 100101, China
zhangjinyang@biols.ac.cn

Agata Muszyńska
Małopolska Centre of Biotechnology, Jagiellonian University, Gronostajowa 7A, 30-387 Krakow, Poland
Institute of Automatic Control, Electronics and Computer Science, Silesian University of Technology, Akademicka 16, 44-100 Gliwice, Poland
a.muszynska@uj.edu.pl

Jeremy Rotman
Department of Clinical Pharmacy, School of Pharmacy, University of Southern California, 1540 Alcazar Street, Los Angeles, CA 90033, USA
rotman@usc.edu

Laura Tao
Department of Computational Medicine, David Geffen School of Medicine at UCLA, 73-235 CHS, Los Angeles, CA 90095, USA
laura.k.tao@gmail.com

Brunilda Balliu
Department of Computational Medicine, David Geffen School of Medicine at UCLA, 73-235 CHS, Los Angeles, CA 90095, USA
bballiu@ucla.edu

Elizabeth Tseng
Pacific Biosciences
1305 O'Brien Drive, Menlo Park, CA 94025, USA
etseng@pacificbiosciences.com

Eleazar Eskin
Department of Computer Science, University of California, Los Angeles, 404 Westwood Plaza, Los Angeles, CA 90095, USA
Department of Human Genetics, David Geffen School of Medicine at UCLA, 695 Charles E. Young Drive South, Box 708822, Los Angeles, CA 90095, USA
Department of Computational Medicine, David Geffen School of Medicine at UCLA, 73-235 CHS, Los Angeles, CA 90095, USA
eeskin@cs.ucla.edu

Fangqing Zhao
Beijing Institutes of Life Science, Chinese Academy of Sciences, Beijing 100101, China
Key Laboratory of Systems Biology, Hangzhou Institute for Advanced Study, University of Chinese Academy of Sciences, Hangzhou 310024, China
zhfq@biols.ac.cn

Pejman Mohammadi
Department of Integrative Structural and Computational Biology, The Scripps Research Institute, 10550 North Torrey Pines Road, La Jolla, CA 92037, USA
Scripps Research Translational Institute, 10550 North Torrey Pines Road, La Jolla, CA 92037, USA
pejman@scripps.edu

Paweł P Łabaj
Małopolska Centre of Biotechnology, Jagiellonian University, Gronostajowa 7A, 30-387 Krakow, Poland
Department of Biotechnology, Boku University Vienna, Muthgasse 18, 1190 Vienna, Austria

pawel.labaj@uj.edu.pl

Serghei Mangul
Department of Clinical Pharmacy, School of Pharmacy, University of Southern California, 1540 Alcazar Street, Los Angeles, CA 90033, USA
serghei.mangul@gmail.com

**Abstract**

RNA-sequencing (RNA-seq) has become an exemplar technology in modern biology and clinical applications over the past decade. It has gained immense popularity in the recent years driven by continuous efforts of the bioinformatics community to develop accurate and scalable computational tools. RNA-seq is a method of analyzing the RNA content of a sample using the modern sequencing platforms. It generates enormous amounts of transcriptomic data in the form of nucleotide sequences, known as reads. RNA-seq analysis enables the probing of genes and corresponding transcripts which is essential for answering important biological questions, such as detecting novel exons, transcripts, gene expressions, and studying alternative splicing structure. However, obtaining meaningful biological signals from raw data using computational methods is challenging due to the limitations of modern sequencing technologies. The need to leverage these technological challenges have pushed the rapid development of many novel computational tools which have evolved and diversified in accordance with technological advancements, leading to the current myriad population of RNA-seq tools. Our review provides a systemic overview of RNA-seq technology and 235 available RNA-seq tools across various domains published from 2008 to 2020, discussing the interdisciplinary nature of bioinformatics involved in RNA sequencing, analysis, and software development.

**Keywords** *RNA-sequencing, Transcriptome quantification, Differential gene expression, High-throughput sequencing, Read alignment, Bioinformatics*

## 1. Introduction

Advances in *high-throughput* sequencing technologies, also known as next-generation sequencing (NGS), have enabled cost-effective probing of gene sequences in living organisms[1]. These sequencing technologies have been adapted to probe the RNA expression (mRNA or totalRNA) of an organism, known as RNA-seq. RNA-seq has reshaped biomedical research by expanding the ability to analyze a vast range of biological data. Biomedical researchers are often tasked with using RNA-seq computational methods which are typically available wrapped as software tools and packages. This review provides an insight into the diverse methodology enabling effective RNA-seq analysis to answer various biological questions, including, but not limited to detecting novel exons, transcripts, gene expressions, and studying alternative splicing structure. We discuss the steps from generating raw data by sequencing technologies to effective interpretation and visualization of this data using mapping, and quantification techniques. Additionally, we performed a comprehensive survey of 235 computational tools designed for various domains of RNA-seq analysis that were developed between 2008 and 2020 **(Figure 1a)**. Our survey assessed the average annual growth rate of computational tools developed for RNA-seq analysis, discusses the most popular tools in each domain of RNA-seq, along with the usability and archival stability of developed computational tools. Encompassing existing knowledge of both the biological and computational foundations involved in RNA sequencing, analysis and software development, we hope this review will lead to a more deliberate use of existing computational tools.

## 2. RNA sequencing

RNA sequencing (RNA-seq) uses *high-throughput* sequencing to determine the order of nucleotides and gene expression. The analysis of RNA-seq data requires specialized computational tools able to address shortcomings of sequencing technologies, including sequencing errors, length biases, and fragmentation. The ability of the computational tools to effectively analyze RNA-seq data has enabled novel therapeutic discoveries, detailed understanding of the regulatory regions that control genes, and identification of biomarkers or mutations in a gene that are known to cause disease onset or progression[2].

RNA-seq begins with preparation of RNA-seq libraries, which are curated collections of RNA fragments where each fragment is sequenced from one or both ends, and the sequence obtained is known as reads. Preparation of an RNA-seq library starts with extracting and isolating RNA from a biological sample, such as a cell line or a frozen tissue sample. The RNA then undergoes reverse transcription and is converted into *cDNA*, which is then amplified by polymerase chain reaction (*PCR*) and fragmented into short sequences (either before or after *PCR*)[3] **(Figure 2).** After the RNA molecules are processed, the RNA-seq library becomes the input for the sequencing machine[4].

### 2.1. Modern high throughput RNA-seq technologies

Modern *high-throughput* sequencing techniques are capable of deriving millions of nucleotide sequences from an individual *transcriptome*. The data obtained from these sequencing technologies provides multifold coverage of the whole *transcriptome* including specific transcriptomic regions that are essential for identifying the underlying causes of common diseases and disorders. High-resolution RNA-seq data can help us identify active genes and quantify the magnitude at which the genes' alternative transcripts are transcribed[5]. The output of these modern sequencing technologies have lengths ranging from hundreds of base pairs (usually referred to as short reads) to thousands of base pairs (i.e., long reads)[1,6,7]. Illumina, Nanopore, and PacBio technologies are among the most commonly used sequencing platforms[8].

The Illumina sequencing platform, based on sequencing-by-synthesis chemistry, was commercialized in 2006. Illumina ligates *cDNA* fragments to adapters and immobilizes the fragments on a solid surface before *PCR* amplification. Next, a reaction mixture containing the primer, reversible nucleotide terminator for each base (labelled with different fluorescent dyes), and the DNA polymerase enzyme, are added to the immobilized solid surface. Each incorporated nucleotide is detected using the *CCD camera* based on the color of the fluorescent dye. The reversible nucleotide terminator and the dye are then removed from the base, and this cycle is repeated until all bases are identified and labelled. The sequences of more than 10 million colonies can be simultaneously determined in parallel, giving rise to a higher sequencing throughput rate when compared to other sequencing platforms[9,10].

Nanopore sequencing technology (including devices MinION, GridIOn, PromethION), introduced in 2014 by Oxford Nanopore Technologies, can produce short or long reads by sequencing native

DNA and RNA fragments of any length. Nanopore technology uses a nanopore, which is a very small hole created by a pore-forming protein or in a synthetic membrane of silicon. The Nanopore sequencing platform simultaneously sends an ionic current and a single strand of DNA or RNA through the nanopore. As the ionic current passes through each nucleotide, it experiences disruptions and variations in a pattern unique to the base pair. These observed patterns of disruption and variation in the current can be interpreted to identify the molecule and for further analysis. While short read sequencing technologies, such as Illumina, require chemical modification or *PCR* amplification, Nanopore technology is capable of sequencing DNA or RNA without these additional steps[12].

The PacBio sequencing technology, commercially known as SMRT (Single-molecule, real-time) Sequencing, was introduced in 2010 and generates full length *cDNA* sequences (i.e., long reads) that characterize transcripts within targeted genes or across entire *transcriptome*s. Long reads generated by PacBio are accurate at the scale of a single molecule through circular consensus sequencing, effectively reading the same *cDNA* insert many times[13,14]. The comparatively high sensitivity of PacBio (Iso-Seq) can be limited by external factors; for example, Iso-Seq has a limitation of being able to produce full-length *cDNA* during the library preparation step. It can only generate high quality reads (HiFi reads) if the target *cDNA* is short enough to be sequenced in multiple passes.

Each sequencing technology exhibits inherent advantages and limitations unique to specific research questions resulting in lack of sequencing technology being suitable for each step of RNA-seq analysis **(Box 1)**. Typically, short read sequencing technologies can generate data with a lower error rate and higher throughput. However, short length of the read makes reconstruction and quantification of the *transcriptome* challenging[15–18]. Long read sequencing technologies improve the accuracy of assembly, or even eliminate the need for assembly (concatenating individual reads to reassemble the *transcriptome*) as each read often covers the entire transcript. Long read sequencing technologies can also be used to produce complete, unambiguous information about alternative splicing events (exons), and improve *genome* annotation to identify the structure, regulatory elements, and coding regions. Long reads currently produce a higher error rate and lower throughput compared to short read technologies **(Figure 3)**[19–21]. Hybrid approaches combining the long and short reads can eliminate the limitations of each platform and can be used to accurately quantify and assemble known and novel transcripts[18,19,22]. Data gathered via these sequencing technologies can address a wide range of research areas including studies that involve *transcriptome* analysis, population-scale analysis, and clinical research[11].

**Box 1. Advantages and limitations of short and long reads**

i. **Error rate** - Short read sequencing technologies have a lower error rate when compared to long read sequencing technologies **(a, b)**.

ii. **Throughput** - The throughput of long read sequencing technologies is typically lower than the throughput of short read sequencing technologies **(c)**.

iii. **Alignment** - Short reads suffer from multi-mapping issues, whereas longer reads, by nature of having more information, can be more accurately mapped to its origin. Due to a high error rate, pairwise alignment between the read, the reference transcriptome, and/or genome is more challenging for long reads compared to short reads.

iv. **Assemble novel transcripts** - Longer reads are preferred for de novo assembly, because they make the assembly step efficient. Most short reads do not span the shared region or shared exon junction, making the assembly step ambiguous. Full-length transcript sequencing eliminates the need for assembly.

v. **Estimate transcripts and gene expression** - Shorter reads are preferred for quantification of transcripts due to their higher throughput. However, assigning short reads to the transcripts requires more advanced probabilistic and statistical approaches. Longer reads have lower throughput, but they can usually cover the entire transcript and make determination of the transcript for each read a straightforward process.

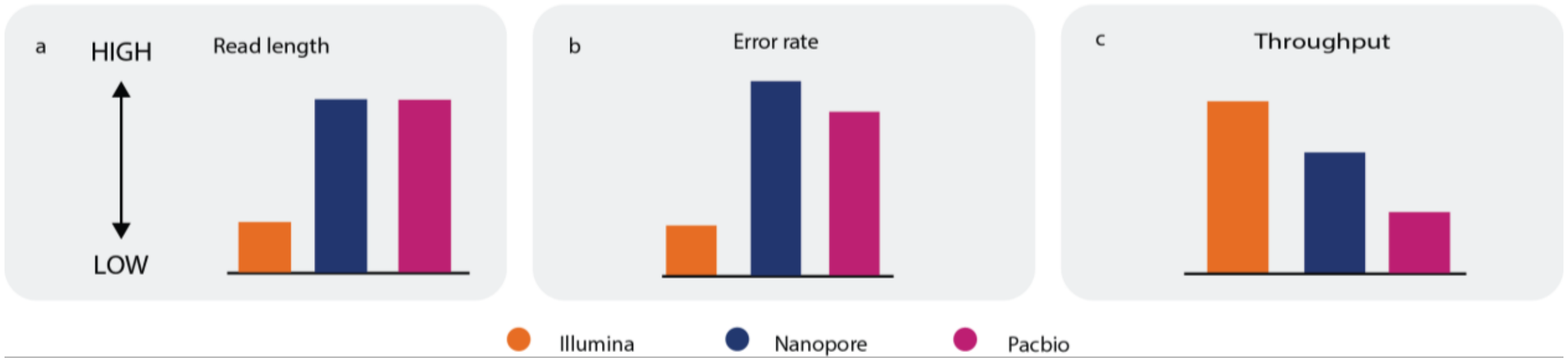

-- End Box

## 3. RNA-seq data science: From raw data to effective interpretation

Analysis of RNA-seq data using computational tools is central to decoding the biological complexities in the human transcriptome. RNA-seq, being multifaceted, can be used to expound and uncover new insights; for example, a dysregulated gene, or a defective protein which could potentially have a downstream effect leading to a diseased state. To illustrate the diversity in RNA-seq data science, we surveyed 235 RNA-seq tools across various domains published between 2008 to 2020. Our analysis indicates that, the average annual growth rate of computational tools developed for RNA-seq analysis was 114.4% from 2008 to 2014 (Figure 1a), but the rate of new tool development slowed after 2015; the average annual growth rate in tools from 2015 to 2020 was 8.97%. These tools, which range across 15 categories in our survey, serve as diverse approaches for the analysis of RNA-seq data (Supplementary Table S.4). Of the 235 RNA-seq tools, most tools were developed for the transcriptome quantification (n=30) (Figure 1b) and the highest number of citations were in the domains of differential expression analysis, followed by data quality control and transcriptome quantification (Figure 1c). On an average the most popular tool in each category accounted for 41.7% of citations.

### 3a. Quality control of raw data

During the sequencing process, potential errors are introduced into reads that can bias the results of downstream analyses such as read mapping and transcriptome assembly. Data quality control and read trimming are essential steps after the reads have been generated, to filter and assess the quality of raw reads[25]. Trimming removes portions of the reads that have lower accuracy. In general, read trimming tools[23,24] aim to increase the accuracy of read alignment and/or transcriptome assembly by removing the bases with a low PHRED score present at the beginning or end of the read. Alternatively, one can apply computational error correction tools to reduce the number of sequencing errors[26,139].

## 4. Read alignment

Read alignment is an essential step in RNA-seq, often required for downstream analysis. The goal is to find the section of the *genome* or *transcriptome* from which the read originated **(Figure 2).** RNA-seq data typically lacks information about the order of the nucleotides and the reads' origin, including the specific part, homolog, or strand of the subject *genome* from which they originate, and this can be established computationally. Additionally, aligning reads to the reference sequence allows for examining the number of reads overlapping each position on the reference sequence, which is known as coverage. Several different bioinformatics tools are capable of estimating the coverage rate and do not require mapping reads to the reference sequence; most of the overlap between reads is preserved with or without the reference sequence[27] **(Figure 2).**

Alignment of the individual RNA-seq reads to their complementary reference sequences can help determine which transcripts are expressed and the strength of an expressed transcript's signal, but the approach is ill-equipped to discover transcripts currently missing from the genome annotation. Assembling the individual *transcriptome* could be challenging and requires increased computational time and resources[28]; instead, known transcripts generated by genome annotation are curated and available from reputable sources such as RefSeq[29], UCSC genome browser,

Ensembl, GENCODE[30] and AceView[31]. However, even the human reference *transcriptome* remains incomplete[32], and unmatched RNA-seq reads should be aligned to the reference *genome* to recover novel transcripts.

One particular computational challenge for aligning RNA-seq reads to the reference sequence is the handling of spliced junctions – one part of the read may map to one end of an exon, while the rest of the read can begin at another exon, often thousands of base pairs apart. Genes create multiple mRNA transcripts through alternative splicing, and, as a result, one or more exons are combined or skipped in different ways and have alternative start/end sites. This varying combination creates multiple transcripts, known as *isoforms*. As a biological process, alternative splicing is evolutionarily advantageous, facilitating efficient cellular reproduction since different protein variants can be generated from the same genetic code **(Figure 3)**. When genome annotations are present, existing exon structures can be used in mapping across known splice junctions; however, this knowledge-guided approach may be biased towards mapping only the known junctions and failing to discover novel ones.

In cases where reads align to multiple transcripts, we may not be able to discern from which *transcript* a read originates. When the *transcript* of origin is unclear for a particular read, our goal is to identify the various alternative splicing arrangements and quantify the transcripts **(Figure 3)**. Splice alignment software packages[33–35] are designed to minimize multi-mapping by correctly aligning these reads across the exon-intron junction of the reference *genome* **(Figure 3)**. Such capabilities can be a crucial first step of reference-guided assembly, wherein transcripts, present in the sample but not annotated in the reference, are assembled using the spliced read alignments to the reference *genome*.

In some instances, reads do not perfectly align with the reference sequence - called mismatches, which can be caused by either sequencing errors or biological variation[26]. RNA-seq alignment tools are typically equipped with a customizable threshold for mismatch or sequencing errors, yet these errors can be due to underlying mutations (genomic variants) in the RNA. Therefore, it is important to distinguish between 'real' deviations from the reference sequence from the sequencing errors. Specialized computational tools[36,37] can identify and classify genes using strategies such as *de novo* assembly (assembly of reads without being aligned to a reference sequence), identification of the reads that span fusion junctions, and filtering, based on many different criteria, of the gene fusion candidates **(Supplementary Table S.4).**

## 5. Quantitative analysis of gene expression levels

The power of RNA-seq allows for quantitative analysis of the gene at the alternative transcript level. As we derive sequence fragments from mRNA, we can detect whether genes are present and how strongly they are expressed, not only at the gene level but also at the transcript level. Additionally, through Differential Expression Analysis (DEA), we can observe whether the strength of the biological signal (i.e., the expression level) is changing and in which direction, allowing for better understanding of mechanisms of living organisms.

### 5.1 Estimation of transcript and gene expressions

Computational methods can effectively leverage the matches between individual reads and reference transcripts by counting the number of reads to estimate expression levels of gene and transcripts. Among all the RNA-seq tools developed between 2008-2020, transcriptome quantification has had the largest number of tools developed, with an average of 3 tools developed per year. The transcriptome quantification tools are among the mostly used tools based on the number of citations **(Figure 1c)**. Counting-based methods[38–40] can estimate gene expression levels by counting aligned reads compatible with gene structure. However, counting-based tools are typically ill-equipped to estimate the expressions of *isoforms* of each expressed *transcript* using the short reads, as the majority of *isoforms* share a large percentage of exons and reads cannot be uniquely assigned to the transcripts **(Figure 3)**. The shorter the reads, the greater the probability that they will match multiple transcripts, presenting a challenge to the identification of the origin of the transcript. A conservative approach to tackle this challenge is to consider only the reads that are uniquely mapped to a single transcript[41]. An alternative approach able to utilize a larger fraction of RNA-seq is to probabilistically assign reads to the *isoforms* from which they originated[42–45].

A number of approaches used in quantifying gene expression do not require read alignment; instead, these tools use raw reads for transcriptome quantification (i.e., reads produced directly by the sequencing machine which are not aligned to the *transcriptome and/or genome*). These ultra-fast methods with small computational footprint can estimate expansion of genes and transcripts using pseudoalignment[46–48]. Pseudoalignment of a read is, a set of target sequences the read is compatible with without any base level alignment. Sailfish[48] pioneered pseudoalignment based on the abundance of the k-mers present in each of the annotated transcripts. Newly developed tools such as Salmon[49] and Kallisto[47] also use pseudoalignment to quantify the *isoforms* of each expressed transcript[46]. A more detailed explanation of these tools can be found in the **Supplementary Note S.2.**

### 5.2 Differential gene expression analysis

After the gene and the transcript expression levels are estimated, statistical approaches are usually employed to detect expression levels of various genomic features (such as genes, exons, transcripts) which exhibit significant statistical differences across experimental groups. In order to perform robust Differential Expression (DE) analysis, first, one needs to reduce *noise* and *false positives* to account for measured sources of expression variation that are unrelated to our variable of interest. For example, *batch effects* such as confounding factors arising from the day the sample was sequenced, laboratory technician, or sequencing laboratory can affect the quantification of the technical replicate of a sample. Existing statistical methods[50–52] can effectively detect and adjust for hidden confounding factors[53].

The most common approach to DEA utilizes statistical testing comparing the strengths of the signal to random change within the population and selecting genes with a multiple hypothesis testing-corrected p-value below a certain threshold, typically 5%. Other approaches in DEA, which may produce more accurate results, use different metrics, such as the Minimum Significant Difference or GLM framework[140], where a combination of p-value with log fold change is applied to identify the most significant genes or alternative transcripts.

An alternative approach is Probability of Positive Log Ratio (PPLR)[141], which was initially developed for microarray analysis and uses a Bayesian hierarchical model to detect DE genes or other genomic features. More recently, it has been adjusted for RNA-seq data[142]. The PPLR expresses the probability or likelihood of the ratio being positive i.e., second condition being upregulated in respect to the first. A PPLR value close to 1 means there is a very high probability of a given *transcript* being upregulated in the second condition. When PPLR is very low, close to 0, this implies very low probability of up-regulation, and consequently the probability of down-regulation being high. There is no direct relation between PPLR and p-value as they come from a different perspective of looking at the problem, i.e., in the probabilistic approach an uncertainty propagation between next stages of analysis is possible and desired. Both of the described approaches might lead to a large number of differentially expressed genomic features. In this case a more stringent p-value cutoff can be applied to reduce the number of DE genes.

Alternatively, depending on the type of *normalization* performed, a machine learning approach can be used wherein, we can choose either classification models based on discrete or continuous distributions. Machine learning approaches have been in use, in the recent past, to manage, model and categorize biological data, enabling high impact discoveries in the field of biomedicine[143]. RNA-seq data is discrete in nature and the two most common approaches[144] are to either model the data as Poisson or negative binomial distribution or second, transform the data to be more similar to microarray data distribution and use already known methods. Bioconductor MLSeq[144] package is a comprehensive source of possible combinations of different *normalization* and machine learning methods. Genes or alternative transcripts (features) can be ranked, or standard sample classification can be performed, and then these features, which mostly contribute to assigning samples to a particular group can be extracted[144]. With a deep learning approach, it is also possible to predict differences in gene expression from histone modification signals[145].

DEA can be complemented by expression quantitative trait *loci* (*eQTL*) analyses, which formally compares the expression levels between different copies (0, 1, or 2) of the minor allele. Each read alignment technique produces different results which may impact which gene will be identified as the DE gene[64]. The power to detect a DE gene and *eQTL*s depends on the sequencing depth of the sample, the minor allele frequency of the genetic variant being tested, the expression level of the gene, and the length of the gene[65]. The magnitude of the *eQTL* can be quantified by log allelic Fold Change (aFC)[76], and its significance is tested using a binomial distribution or the over dispersed generalizations[77–81].

The results of these transcriptomic analyses can be validated using independent techniques such as *qPCR*, which is statistically assessable[69]. The measurement of gene expression obtained by *qPCR* is relatively similar to the measurement obtained by RNA-seq analysis, where a value can be calculated for the concentration of a target region in a given sample[70–72]. Additional information about the Quantification of RNA splicing and sQTL analyses can be found in **Supplementary Note S.3.** Among the surveyed RNA-seq tools, 24 were DEA tools[66,67], especially ones representing a classical, statistic-based approach, including DESeq2[68] which was cited the most number of times making it one of the most popular RNA-seq tools. The DEA tools also had the largest number of citations per year **(Supplementary Fig**. **S.6)**.

## 6. Measuring allele specific expression from RNA-seq

RNA-seq can also be used to study allele specific expression (ASE or allelic expression) to study the cis-regulatory effects of genetic variants[64–66]. ASE represents gene expression independently measured for the paternal and maternal allele of a gene. In a typical RNA-seq experiment, ASE can be measured only in genes where the sequenced individual is carrying a heterozygous *SNP* within the transcribed region. This *SNP*, referred to as the aseSNP, can be used as a tag to identify RNA-seq reads that originate from each of the gene alleles and produce ASE data **(Figure 4).**

The allelic imbalance—the ratio between paternal and maternal allele gene expressions—is, perhaps, the most interesting signal in ASE data and identifies genetic cis-regulatory differences between the two haplotypes. aFC[76] can also quantify the magnitude of allelic imbalance. An aseSNP is not itself a regulatory variant and should not induce an imbalanced ASE signal. However, there can be a bias in ASE data that falsely implies the haplotype carrying the reference allele for the aseSNP has a slightly higher expression across all genes. This issue, known as allelic bias or reference bias, can be mitigated by aligning reads to personalized reference *genomes*, excluding likely biased sites[33,73,82–84] or aggregating the ASE signal from multiple aseSNPs in each gene[85]. ASE data can also be used to improve statistical power for identifying *eQTL*[78, 82–84], and fine mapping the causal regulatory variants in *eQTL* data[88–90]. ASE data is inherently robust to noise; thus, it can also be used to identify gene environment interaction effects[91], or to identify the effect of rare genetic variants on gene expression in order to improve diagnostic accuracy for Mendelian diseases[92,93]. Tools and best practices[73] for generating and analyzing ASE data are discussed in more detail in **Supplementary Table S.4**.

## 7. Profiling circular RNA from RNA-seq

Circular RNA (circRNA) is a large class of RNA molecules with covalently closed circular structure that plays important roles in various biological processes and metabolic mechanisms[94]. In recent years, a variety of computational tools have been developed for circRNA study[84,85]**(Supplementary Table S.4)**. As an initial step in circRNA analysis, identification of circRNAs is currently based on detection of reads spanning the circle junction, which is termed the back-splice junction (BSJ). Most tools[86–88] employ aligners[100–102] to detect putative back-spliced events from fusion-reads or split alignment results, while others splice-aware aligners[35,103] can align circular reads and detect back-splice junctions directly.

Considering that most circRNAs are derived from exonic regions[93,94] where computational methods are unable to accurately distinguish linear and circular reads, BSJ read count becomes the most reliable measurement of circRNA expression levels. For count-based methods, BSJ read count is inferred from alignment results, and different filters and statistical strategies have been employed to improve the accuracy and sensitivity of these methods[98,106]. Similar to the quantification of linear transcripts, pseudoalignment-based tools for circRNA quantification[96] substantially increased the computational efficiency compared to regular alignment-based methods. To effectively compare the expression levels of circRNAs and their host genes, the junction ratio, defined as the ratio of BSJ reads and linear reads mapped to BSJ site, is often used for comparative analysis. Several computational methods have been developed for accurate estimation of junction ratios[108,109]. Moreover, circRNAs also exhibit alternative splicing patterns, and a number of specific tools have been developed for circular transcript assembly[86,99–101], internal structure visualization[113,114] and differential expression analysis[98,109]. Additionally, several comprehensive databases have been constructed for circRNA annotation and prioritization analysis[94,104,105].

## 8. Usability and archival stability of RNA-seq tools

Maintaining the archival stability of bioinformatics tools is increasingly important in preserving scientific transparency and reproducibility. We assessed the archival stability of 235 RNA-seq tools. The majority of RNA-seq tools are stored on archivally stable repositories (e.g, GitHub) and other tools are hosted on personal or academic webpages **(Figure 5b)**, which often have limited archival stability [117].

We have also assessed the computational expertise required to install and use RNA-seq tools. A vast majority of tools require the user to operate the command line interface **(Figure 5a)**. Web-based tools accounted for 8.09% of tools. There has been an increasing number of tools with available web-based interfaces designed for small-RNA detection and cell deconvolution categories (71.4% and 31.2%, respectively). We have also compared the availability of package managers across RNA-seq tools. Package managers are platforms that automate the installation, configuration, and maintenance of a software tool, promising to expedite and simplify the installation of a software tool and any required dependencies. The majority of RNA-seq tools lack a package manager implementation **(Figure 5c and 5d).** For the tools with available package manager implementation, Anaconda[118] was the most commonly used package manager platform. The second most popular platform was Bioconductor[119] and CRAN[120].

Lastly, we evaluated the effect of usability on the popularity of RNA-seq tools. We found that tools that are available as package managers had significantly more citations per year compared with tools which are not available as package managers **(Figure 5e;** Mann–Whitney U test, p-value = $1.85 \times 10^{-7}$).

## 9. Discussion

As technology has advanced, RNA-seq methods have become increasingly popular and have revolutionized modern biology and clinical applications, driven by continuous efforts of the bioinformatics community to develop accurate and scalable computational tools. In addition, NGS technologies have provided the unprecedented ability to analyze a wide range of biological data, allowing for the exploration of both novel and existing biological problems. On an average, across the domains, there have been 18 tools developed each year between 2008-2020. To address this barrier, we provide an overview highlighting the fundamentals of RNA-sequencing and its associated computational methods. In this review, we provide an overview of modern sequencing technologies and discuss the advantages and limitations associated with each type.

RNA-seq data analyzed by computational tools can be used to effectively tackle important biological problems such as estimating gene expression profiles across various phenotypes and conditions or detecting novel alternative splicing on specific exons. Specialized analyses of RNA-seq data can also help detect changes in concentration, function, or localization of transcription factors that affect splicing and can cause onset of neurodegenerative diseases and cancers[121,122]. More recently developed computational tools[112–116] are capable of repurposing RNA-seq data characterizing the individual adaptive immune repertoire and microbiome[128]. Additionally, RNA-seq data can also be used to study cell type compositions using computational deconvolution, a process of estimating cell type proportions in tissue samples[129,119].

This interdisciplinary nature of bioinformatics involved in RNA sequencing, analysis, and software development introduces undefined terms that challenge novice researchers in the wider scientific and medical research community. The current literature on RNA-seq methods have traditionally assumed that the reader is familiar with fundamental concepts of RNA-seq technologies and the implied bioinformatics analyses[131–136]. These methods also require diverse computational skills to be used effectively. The lack of computational skills among the biomedical researchers limit the ability to unlock the potential of RNA-seq analysis; therefore requiring a review that explains rudimentary concepts and defines discipline-specific jargon.

In addition to an overview on the fundamentals of RNA-seq, our review also includes survey results of 235 computational tools developed from 2008 to 2020 for various domains of RNA-seq analysis are available in this easy-to-use resource - https://github.com/Mangul-Lab-USC/RNA-seq. In addition to information about usability and archival stability of the tools, this resource will engage the biomedical community through sharing the feedback on utilizing the tools.  We hope our resources will help researchers make a more informed decision when selecting a tool for a specific type of data and research question.

## Methods

### Determine features of RNA-seq tools

We compiled 235 RNA-seq tools published between 2008 and 2020, which have varying purposes and capabilities based on the type of analysis one is conducting or the biological questions one is answering.

We identified 15 primary areas of application in RNA-seq analysis (Data quality control, read alignment, gene annotation, transcriptome assembly, transcriptome quantification, differential expression, RNA splicing, cell deconvolution, immune repertoire profiling, allele specific expression, viral detection, fusion detection, small RNA detection, detecting circRNA, and visualization tools). After assigning each tool a category based on its area of application, we highlighted the "Notable Features" for each tool. These notable features encompassed a range of functionalities: purpose of the tool; features that render the tool unique or not unique within its category; the form of receiving RNA-seq data as an input; the form of presenting output.

We documented whether each tool was web-based or required of the user one or many programming languages for installation and/or utilization ("Programming Language"). In addition to the programming languages, we highlighted whether a package manager (e.g., Anaconda, Bioconductor, or CRAN) was available. Based on the combination of which programming language was required and which package manager was available for each tool, we assessed the required expertise needed to be able to install or run the tool. If the tool was a web-based tool with no package manager available or a web-based tool along with programming languages and package manager present, we assigned the tool the little-to-none required expertise of "+". If the tool required only R as a programming language or along with other programming languages and had a package manager present, the tool was assigned a required expertise of "++". In addition, if programming languages other than R were required, and a package manager was present, the tool was assigned a "++". Lastly, tools that required a single programming language (other than R) or multiple programming languages, and lacked a package manager to aid installation, were assigned a "+++" for the most required expertise.

Each published tool had a designated software link where the tool can be downloaded and installed. Based on the type of platform hosting the URL, we assigned the tools a "1" or "2". An assignment of "1" meant that the tool's software was hosted on a more archivally stable web service designed to host source code. An assignment of "2" meant that the tool's software was hosted on a less archivally stable web service (e.g., personal and/or university web services).

**Code and data availability**

https://github.com/Mangul-Lab-USC/RNA-seq

**Acknowledgements**

P. Mohammadi's work has been supported by NIH CTSA grant (UL1TR002550). Fangqing Zhao's work has been supported by from National Natural Science Foundation of China (31722031, 91640117). Agata M's work has been co-funded by the European Union through the European Social Fund grant (POWR.03.02.00-00-I029) We thank Dr. Patro for his valuable comments and suggestions on the manuscript.

**Competing Interests**

The authors declare no competing interests.

## Figures and figure legends

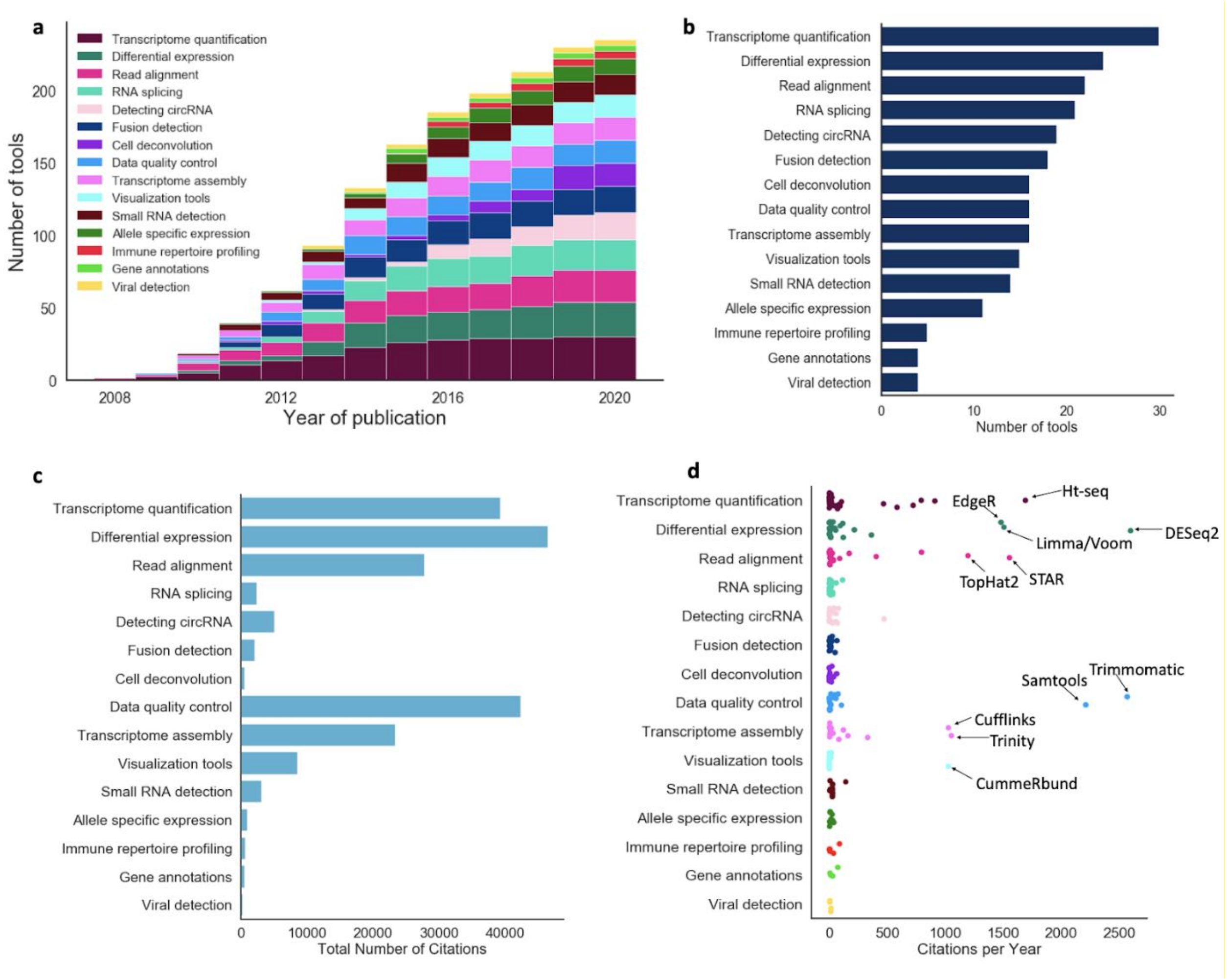

**Figure 1: Landscape of RNA-seq tools** (a) The cumulative number of current computational tools for RNA-seq analysis from 2008 to 2020. (b) The number of current computational tools for RNA-seq analysis in each category. (c) The total number of citations in each domain from 2008-2020. (d) The popularities of tools in different categories, determined by citations per year since initial release. The most popular tools in some of the categories are labeled.

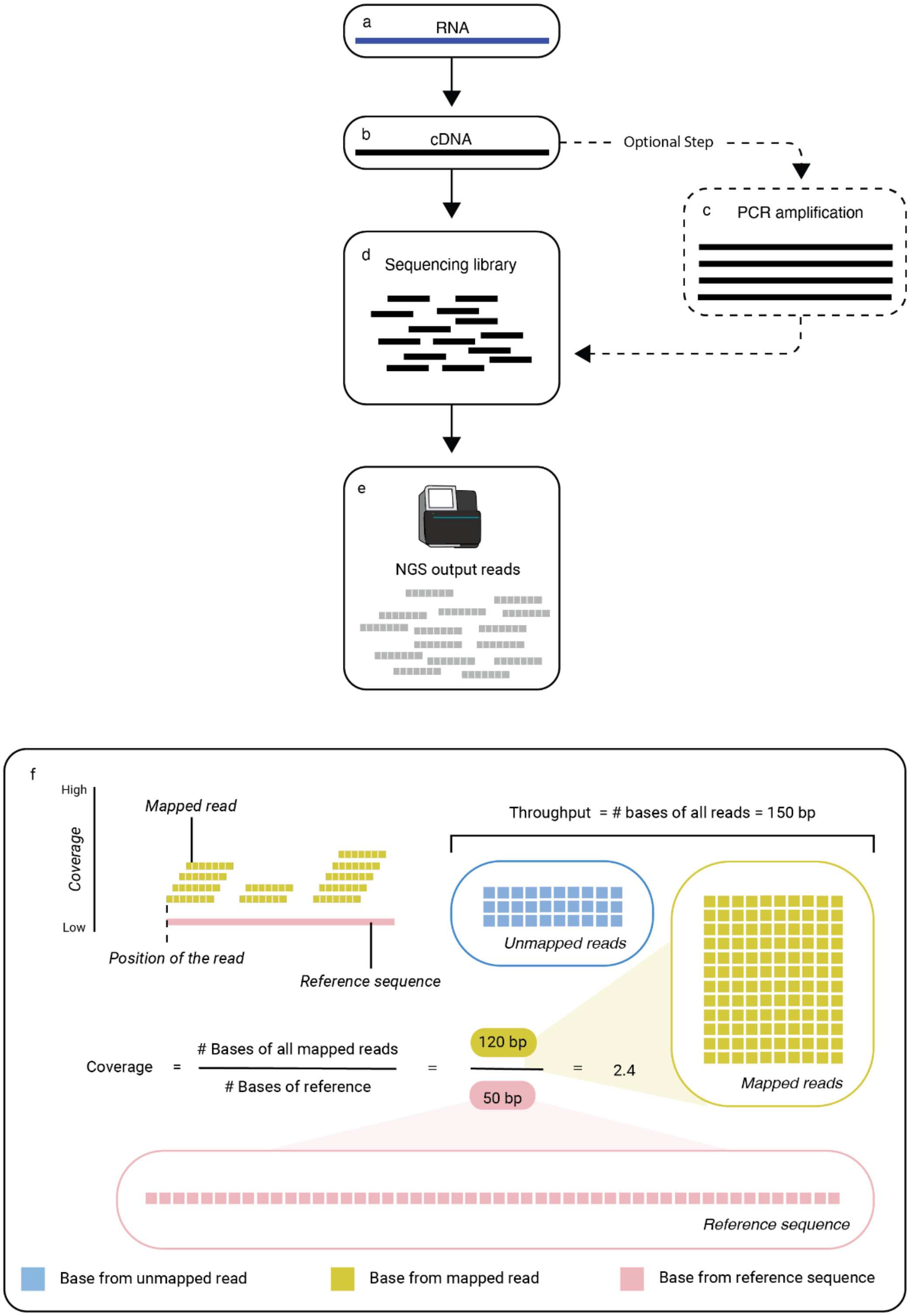
a
RNA
b
cDNA
Optional Step
c
PCR amplification
d
Sequencing library
e
NGS output reads
f
High
Coverage
Low
Mapped read
Position of the read
Reference sequence
Throughput = # bases of all reads = 150 bp
Unmapped reads
Mapped reads
Coverage = # Bases of all mapped reads / # Bases of reference = 120 bp / 50 bp = 2.4
Reference sequence
Base from unmapped read
Base from mapped read
Base from reference sequence

**Figure 2. Overview of RNA-seq**

RNA-seq is a process of creating short sequencing reads from RNA molecules. The steps consist of first converting the RNA (a) into cDNA (b) and then amplifying the cDNA by PCR (c - which is an optional step) and lastly fragmenting the cDNA into short pieces (known as fragments). After the sequencing library (d) is prepared, the fragments are used as input for the Next Generation Sequencer (e). The Next Generation Sequencer then sequences each fragment generating FASTQ files. (f) In read alignment, reads from a sequencing machine are provided. The modern high throughput sequencing machines are able to generate up to 150 million reads. The reference sequence, shown as a pink line, is known. The goal of alignment is to find the loci in the reference genome that mostly matches a read sequence. A read is shown to align to the position x1, x2, x3 and the position is recorded.

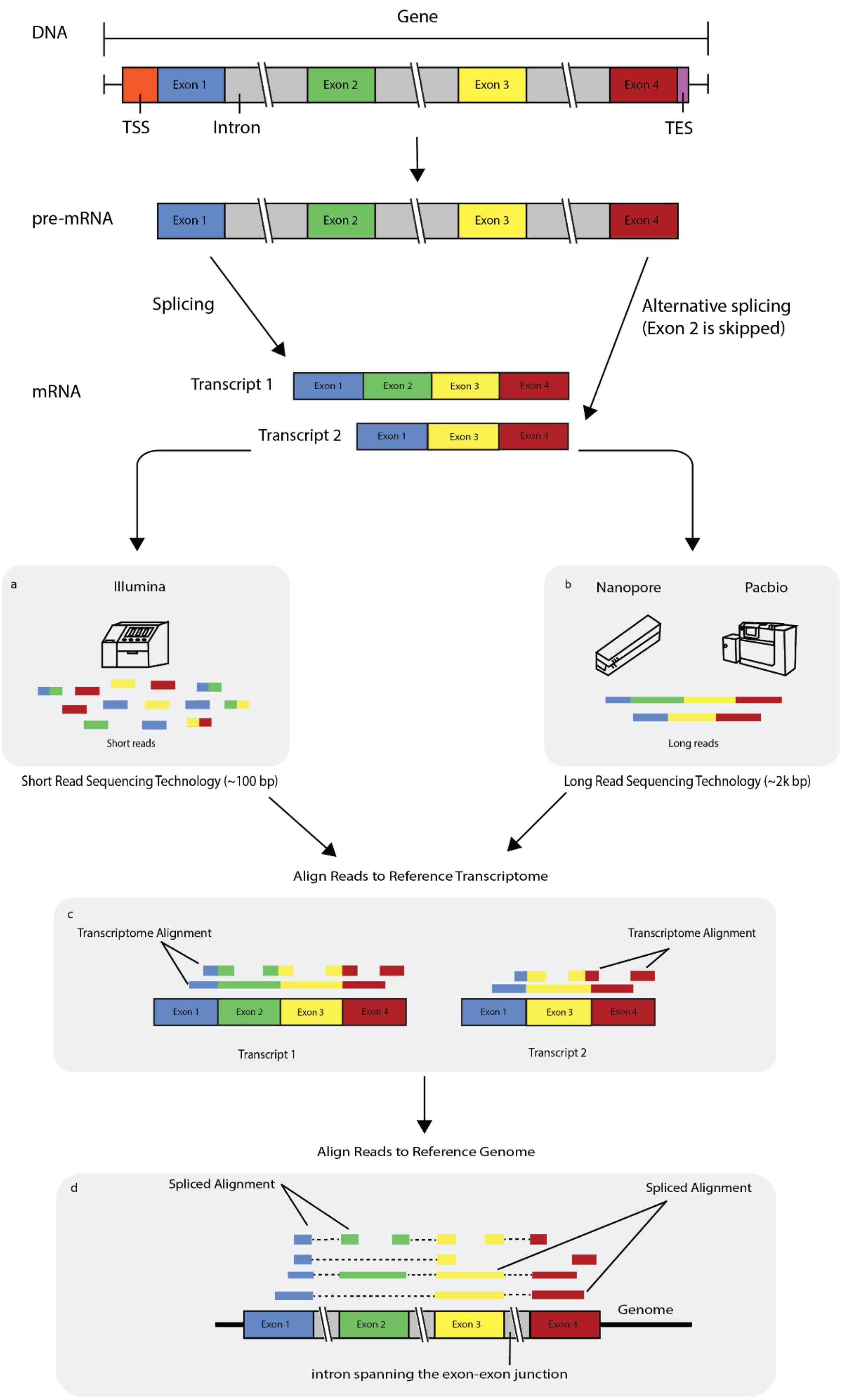

**Figure 3: Alternative splicing and RNA-Seq technologies.**

The flow of genetic information begins with the DNA, which consists of introns and exons. DNA is transcribed into pre-mRNA, further processed into mature mRNA, splicing out the introns and leaving the exons glued together. mRNA is further translated into a protein.

Different arrangements of exons (transcripts) may be formed in a process called alternative splicing or exon skipping. An RNA-seq read is a short sequence sampled from a transcript. These reads are generated using modern sequencing technologies such as (a) the Illumina platform - a short-read sequencing technology, (b) Nanopore and PacBio platforms - long-read sequencing technologies. The figure depicts two scenarios which illustrate the alignment of the uniquely mapped reads to the transcriptome (c) and the genome (d). A few of the reads are multicolored indicating that, when aligned, they span across the exon-exon junction. Some of the shorter reads (single-colored) are aligned only to a single exon and do not span across the junction.

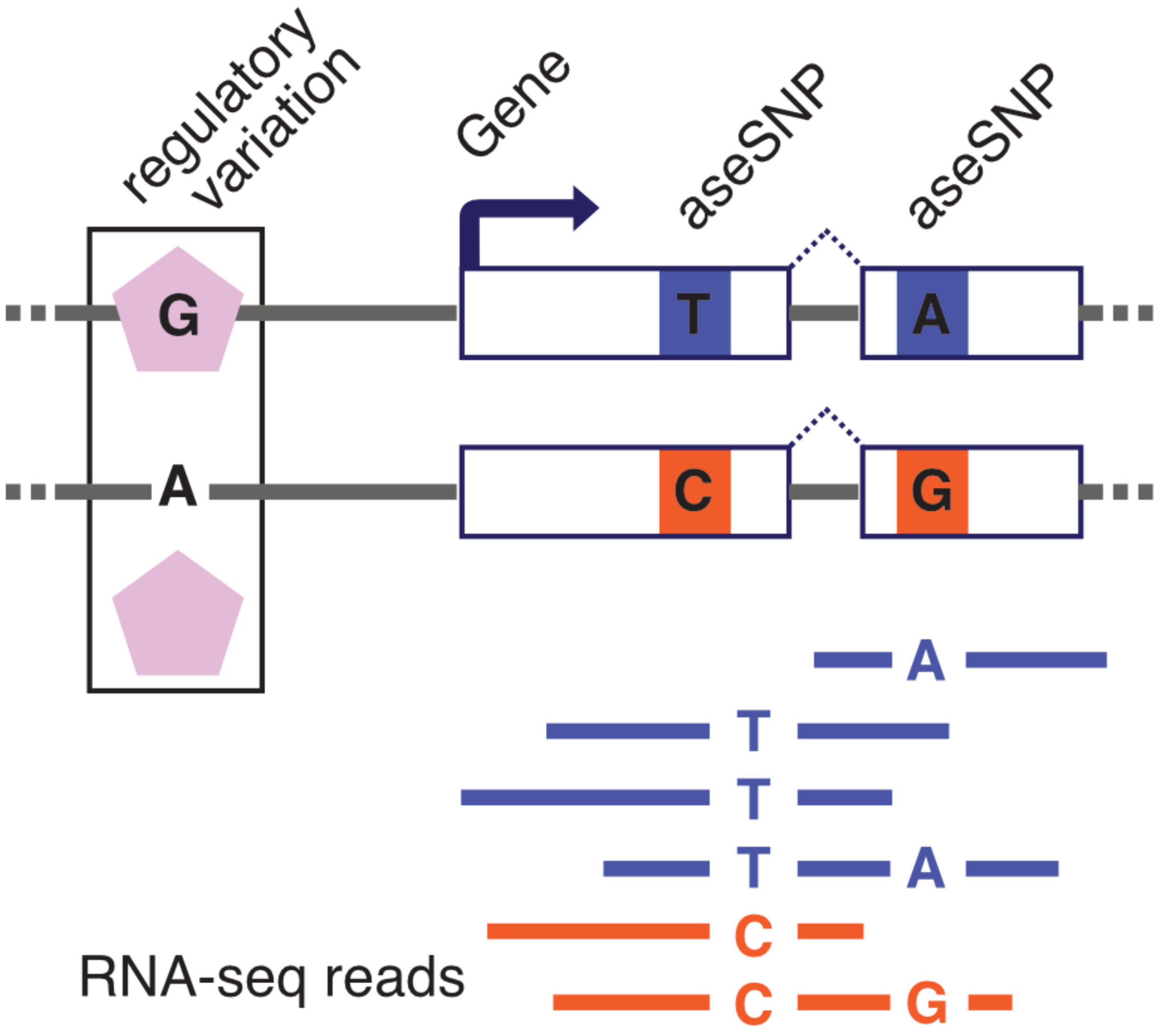

**Figure 4: Measuring allele specific expression from RNA-seq.**

RNA sequencing can be used for generating ASE data for genes in which an individual is carrying a heterozygous SNP in the transcribed region (aseSNP). aseSNP alleles allow for mapping sequencing reads back to the haplotype the originate from. Imbalance in ASE data is a functional indicator of cis-regulatory difference between the two haplotypes that is driven by heterozygous regulatory variants. Data from multiple aseSNPs can be aggregated to improve ASE data quality.

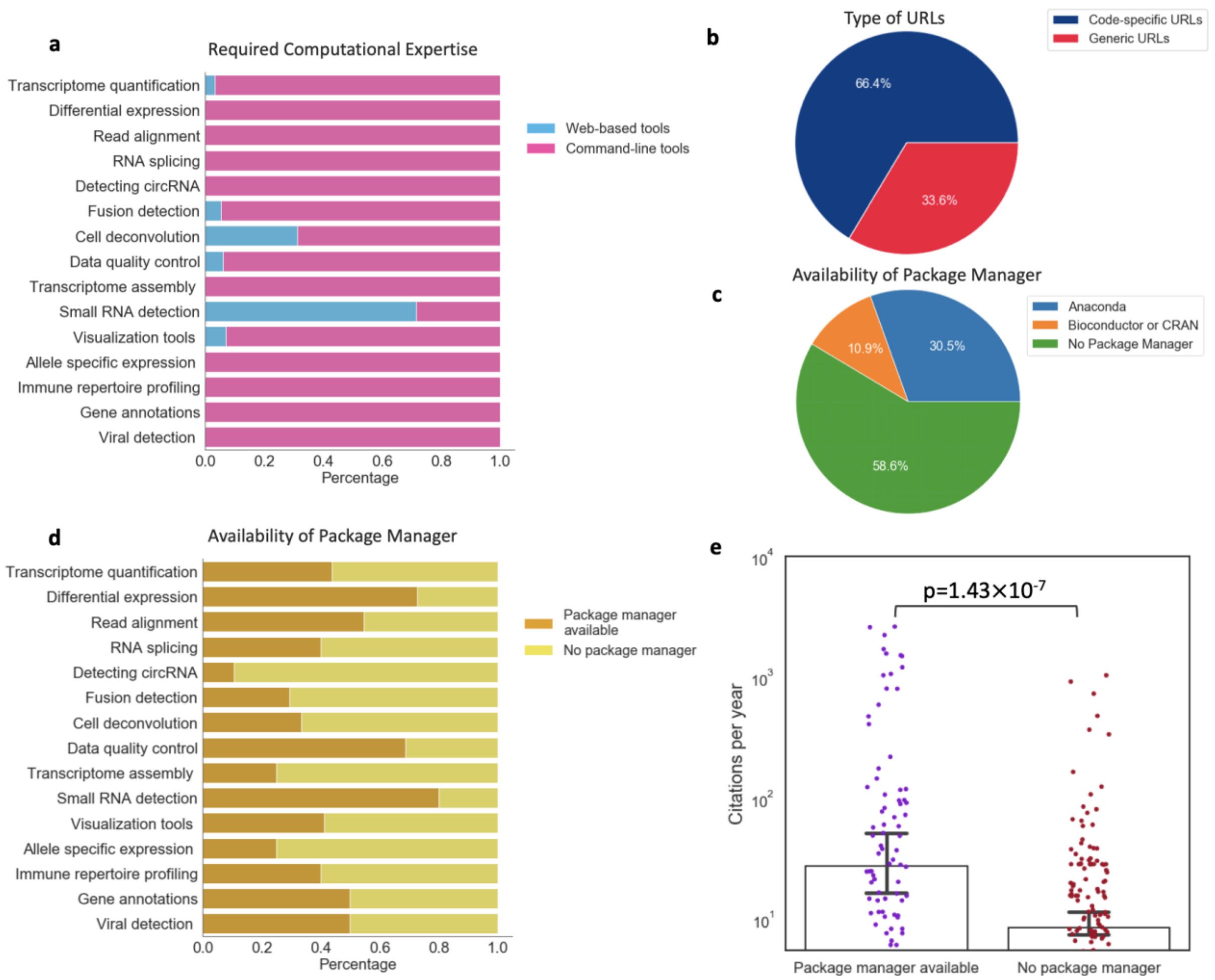

**Figure 5. Usability and archival stability of RNA-seq tools analysis** (a) The percentage of required computational expertise of current computational tools that are used for RNA-seq analysis in different categories. (b) The percentages of type of URLs for tools collected. (c) The percentages of package managers for all tools collected. (d) The percentage of availability of package managers for tools in different categories. (e) Tools that are available as package managers exhibit increased citations per year compared with tools that are not available as package managers (Mann–Whitney U test, p-value = $1.85 \times 10^{-7}$).

**Supplementary Materials**

**Glossary**

1. High throughput: High throughput refers to the sequencing methods which are capable of or enable the sequencing of multiple DNA molecules (thousands of strands) at a time.
2. cDNA: complementary DNA or cDNA is the DNA produced by reverse transcription of the template RNA strand by the action of the enzyme reverse transcriptase. The cDNA being synthesized remains single stranded (mRNA) until PCR is carried out. In PCR, aided by the DNA polymerase, it gets its double stranded form.
3. Genome: Genome is the collection of all the DNA present in the nucleus and mitochondria of the cell.
4. Transcriptome: Transcriptome is the collection of all the mRNA molecules derived from or expressed in the genes of an organism.
5. Transcript: A transcript is the single stranded RNA molecule produced as a result of transcription (copying of a particular region of the DNA into RNA by an enzyme RNA polymerase)
6. Isoform: Isoforms are transcripts produced from the same gene but have different TSS and potentially different functions as a result.
7. Loci: Locus (plural - loci) is a specific location or position on a gene.
8. PCR: Polymerase chain reaction is a method used to amplify or create multiple copies of a specific DNA sample rapidly.
9. Nucleotide fluorescence detection: Nucleotide fluorescence detection is a method of fluorescently labelling the nucleotide bases for detection of each base and deducing the nucleotide sequence.

10. Phred quality scores: Phred quality score is a measure of accuracy of sequencing or the estimate of probability of an error in the base call; also called Q score. This score is provided by the NGS machine.
11. CCD camera: Charge-coupled device camera used for digital imaging in sequencing.
12. Poly-A tails: Poly-A tail is a long chain of Adenine nucleotides added to mRNA for protection from enzymatic degradation, increasing its stability.
13. qPCR: quantitative PCR measures the amplification of the reverse transcribed DNA code. It is used for the quantification and analysis of nucleic acids for numerous applications.
14. eQTL: expression quantitative trait loci focusing on amount and not the type of trait.
15. Batch effect: Batch effect is the difference that can be observed within the technical replicates of a sample which could be due to various factors such as the day or place where it was sequenced, the researcher conducting the experiment, etc.
16. Normalization: Normalization is the process of adjusting or removing confounders and variations in the data caused due to batch effect.
17. SNP: A single nucleotide polymorphism is the replacement of a single nucleotide (base pair) with another nucleotide at a particular position in the genome.

## S.1 Supplementary Note 1

### Alternative Splicing:

The presence of alternative splicing in RNA presents a challenge to aligning reads to a reference genome for RNA-seq reads. Alternative splicing begins during transcription, when DNA is used as a template to create a single strand of RNA. Typically for protein coding transcripts, first, the DNA molecule is copied by RNA polymerase into a form known as pre-mRNA, which contains all the bases of the original DNA strand. Next, portions of the pre-mRNA, called introns, are removed and the remaining portions, called exons, are combined. The final mRNA referred to as a transcript, is ultimately translated into a protein, or has other functions as a noncoding RNA (ncRNA). Introns are excised from the RNA transcripts, but the order of exons that comprise these transcripts vary. Accounting for this loss of sequence information from the intron requires that we look to the junction regions of where any two exons are combined **(Figure 3).**

## S.2 Supplementary Note 2

### Sailfish, Salmon and Kallisto

Sailfish[48] pioneered pseudoalignment, based on the abundance of the k-mers present in each of the annotated transcripts. This reformulation replaces the problem of read alignment with the method of k-mer counting, which is considerably faster. K-mers are substrings of length ‘k’ in a nucleotide sequence. For example, a dinucleotide sequence is a k-mer, where k = 2. In the sequence CCTAGTGTACCGTACC, CC is a 2-mer with a frequency of three. This approach represents a reduction of the EM algorithm from modeling the set of observed fragments, as is done in [43] to modeling the set of fragment counts; such a reformulation was also introduced by [50] at the per-gene level and by [51] at a transcriptome-wide level, where the equivalence classes are defined over RNA-seq fragments rather than k-mers. To further accelerate EM convergence, Sailfish makes use of the SQUAREM algorithm [52]. While vastly accelerating quantification, this approach neglects accounting for coherence between the k-mers within a read or read pair, which can diminish accuracy.

Most recent methods also leverage raw reads for quantifying gene expression, are computationally efficient, and use lesser memory than alignment-based approaches comparable to methods [42,46,47]. Kallisto's algorithm deciphers the compatibility of reads with transcripts but does not map the entire read [53]. This pseudoalignment approach increases the speed of computation and decreases the memory requirement. Each k-mer is matched with a group of transcripts, and the intersection of the sets of transcripts represent the best possible matches[54]. These transcript compatibility sets define equivalence classes of fragments, and the equivalence classes and counts are used as sufficient statistics, as in [51], to estimate transcript-level abundances using an EM algorithm.

Salmon offers two different modes for quantification of transcripts. One mode includes both mapping and quantification in a single step, with reads being mapped directly to the transcriptome using the quasi-mapping approach of RapMap[55], without creating any intermediate alignment files [49]. Alternatively, transcriptome-aligned SAM or BAM files — such as those used by RSEM [43]— can be provided. Salmon leverages a two-phase inference algorithm for quantification. The first phase consists of a stochastic Bayesian inference procedure that models the reads at fragment resolution (which can be used to estimate certain effects, like fragment GC-bias or alignment error profiles, that depend on the characteristics of individual read mappings), while the second phase operates on equivalence classes of fragments, either as in[51] or using refined fragment equivalence notions that have been introduced subsequently[56], and estimates abundance using either an EM or variational Bayesian EM [57] algorithm.

## S.3 Supplementary Note 3

### Quantification of RNA splicing and sQTL analyses

Computational tools that capture alternative splicing events in RNA seq data can characterize the possible set of transcripts within the data that were missing from multi-mapped reads. Further insight regarding differentially expressed genes can then be obtained by comparing this broader set of transcripts between case and control groups.

Estimation of isoform ratios or exon inclusion levels is currently the standard approach to studying splicing events in RNA-seq data. Direct inference based on the excised introns is the second most common approach to study splicing events. Accurate quantification of splicing events can be obtained using methods such as LeafCutter [137] or MAJIQ-SPEL [138]. MAJIQ-SPEL [138] is often used together with VOILA (a visualization package) to define, quantify, and visualize local splicing variations from RNA-seq data. LeafCutter [137] quantifies RNA splicing variations using short read RNA-seq data; in other words, the program anchors reads that span an intron to quantify the intron usage across samples. By detecting novel introns, LeafCutter [137] can help researchers avoid the additional burden of performing isoform abundance estimation [137,139].

Once the ratios of each isoform, exon, or excised intron are available, splicing levels across different treatment groups or different copies of the minor allele can be compared with differential splicing analyses and splicing QTL (sQTL) analyses. LeafCutter [137] can detect differential splicing between sample groups and map sQTLs. ulfasQTL [140] and sQTLseekeR [141] are commonly selected for mapping sQTLs.

**Supplementary Table**

**S.4 Table**

<attached as a separate document>

**Supplementary Figures**

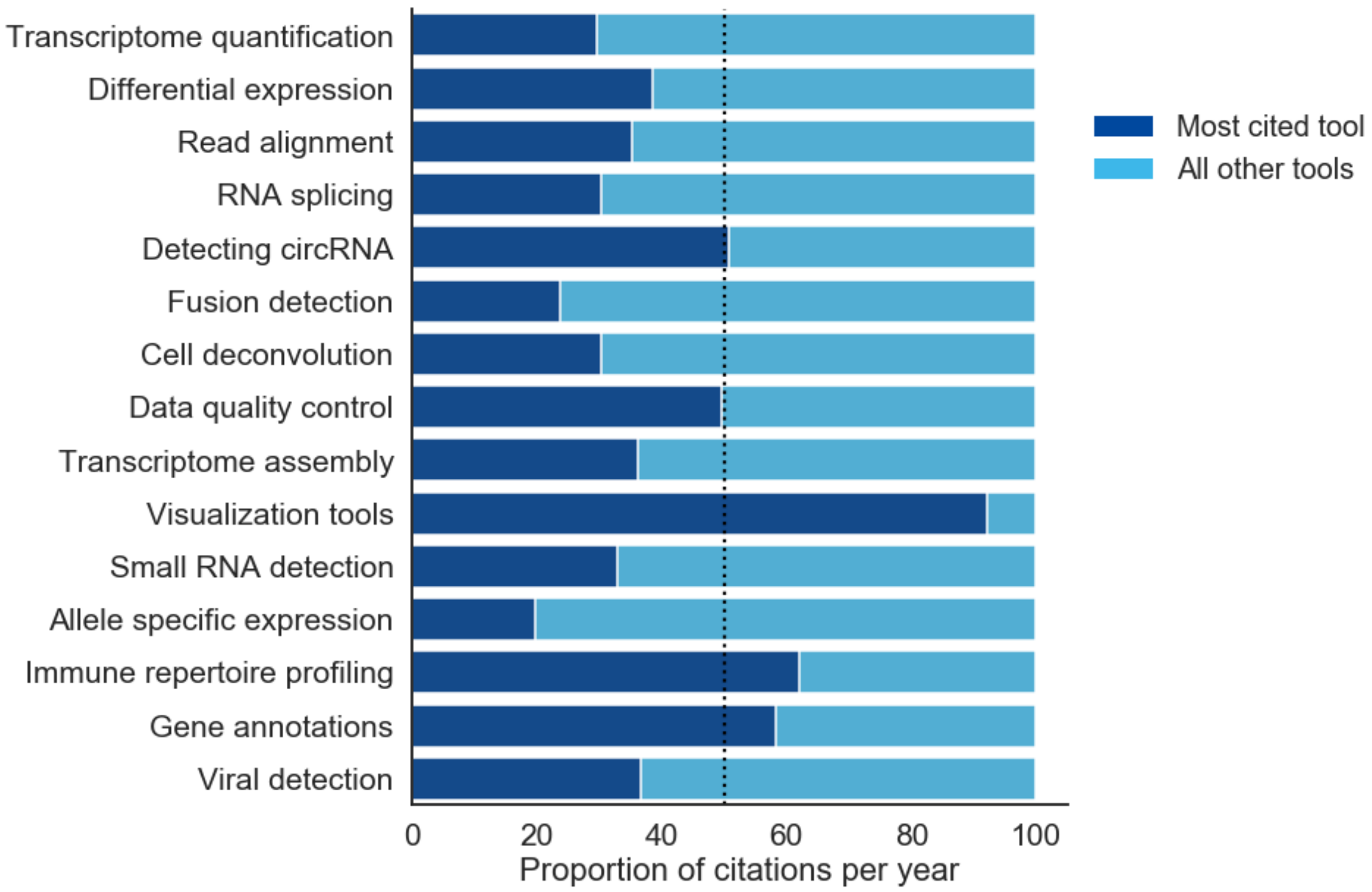

**S.5 Proportion of citations of RNA-seq tools per year.**

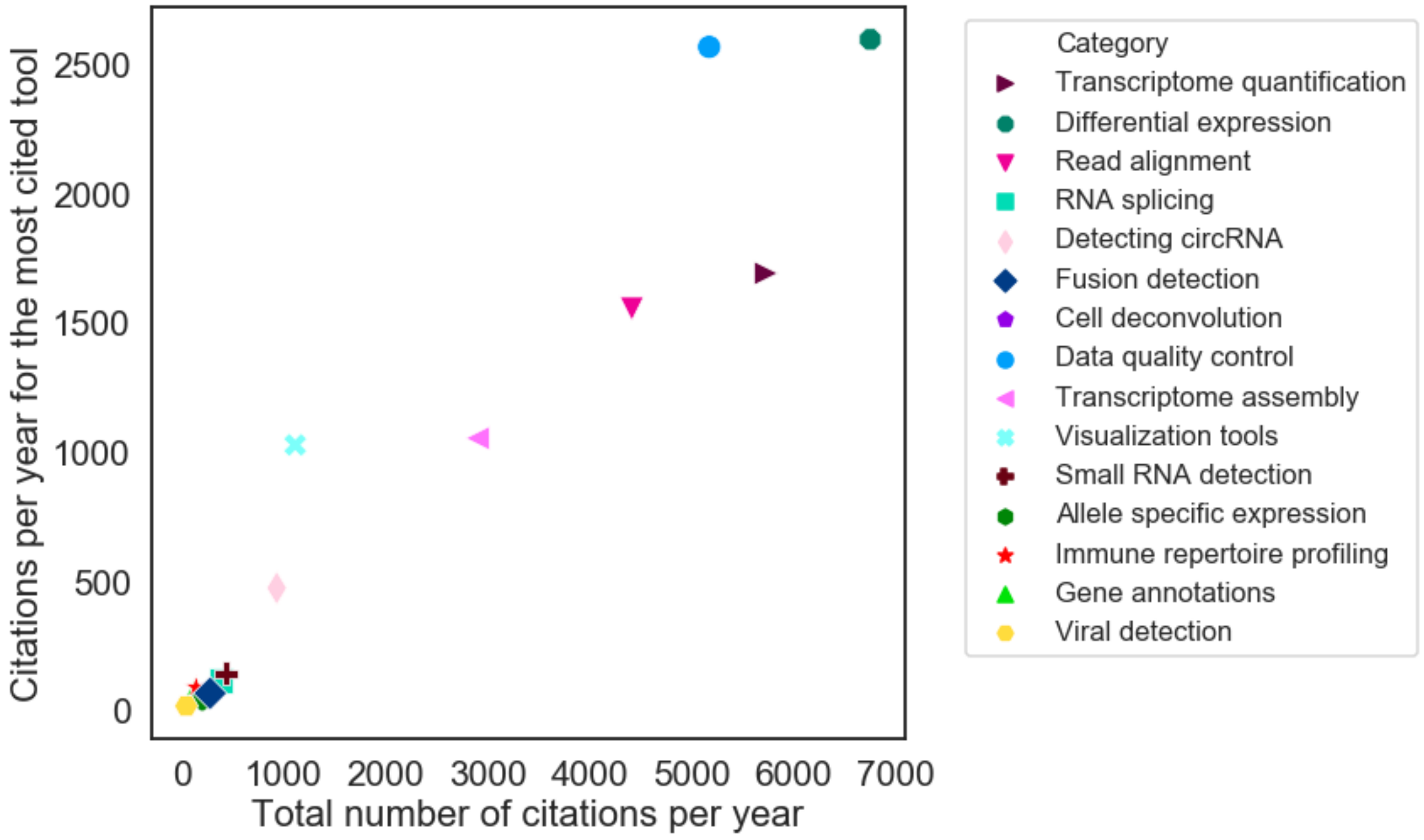

**S.6 The relationship between total number of citations per year of each category and the number of citations per year for the most cited tool in that category.**

**S.4 Table**. Landscape of current computational tools for RNA-seq analysis

| Tool | Year Published | Notable Features | Programming language | Package manager | Required expertise | Software | Type of URL<br><br>1. Web services designed to host source code<br><br>2. Others (e.g personal and/or university web services) |
|---|---|---|---|---|---|---|---|
| **a. Data quality control** | | | | | | | |
| iSeqQC[1] | 2020 | Expression-based raw data QC tool that detects outliers | R | N/A | ++ | https://github.com/gkumar09/iSeqQC | 1 |
| qsmooth[2] | 2018 | Adaptive smooth quantile normalization | R | Bioconductor | ++ | http://bioconductor.org/packages/release/bioc/html/qsmooth.html | 1 |
| FastQC[3] | 2018 | Raw data QC tool for for high throughput sequence data | Java | Anaconda | ++ | https://github.com/s-andrews/fastqc/ | 1 |
| QC3[4] | 2014 | Raw data QC tool detecting batch effect and cross contamination | Perl, R | Anaconda | ++ | https://github.com/slzhao/QC3 | 1 |
| kPAL[5] | 2014 | Alignment-free assessment raw data QC tool by analyzing k-mer frequencies | Python | Anaconda | ++ | https://github.com/LUMC/kPAL | 1 |
| HTQC[6] | 2013 | Raw data QC read assessment and filtration | C++ | N/A | +++ | https://sourceforge.net/projects/htqc/ | 1 |
| Trimmomatic[7] | 2014 | Trimming of reads and removal of adapters | Java | Anaconda | ++ | http://www.usadellab.org/cms/index.php?page=trimmomatic | 2 |
| Skewer[8] | 2014 | Adapter trimming of reads | C++ | Anaconda | ++ | https://sourceforge.net/projects/skewer | 1 |
| Flexbar[9] | 2012 | Trimming of reads and adaptor removal | C++ | Anaconda | ++ | https://github.com/seqan/flexbar | 1 |
| QuaCRS[10] | 2014 | Post QC tool by performing meta-analyses on QC metrics across large | Python | N/A | +++ | https://github.com/kwkroll32/QuaCRS | 1 |
| | | numbers of samples. | | | | | |
| BlackOPs[11] | 2013 | Post QC tool that simulates experimental RNA-seq derived from the reference genome and aligns these sequences and outputs a blacklist of positions and alleles caused by mismapping | Perl | N/A | +++ | https://sourceforge.net/projects/rnaseqvariantbl/ | 1 |
| RSeQC[12] | 2012 | Post QC evaluation of different aspects of RNA-seq experiments, such as sequence quality, GC bias, nucleotide composition bias, sequencing depth, strand specificity, coverage uniformity and read distribution | Python, C | Anaconda | ++ | http://rseqc.sourceforge.net/ | 1 |
| | | over the genome structure. | | | | | |
| RNA-SeQC[13] | 2012 | RNA-seq metrics for post- quality control and process optimization | Java | Anaconda | ++ | https://software.broadinstitute.org/cancer/cga/rna-seqc | 2 |
| Seqbias[14] | 2012 | Post QC tool using a graphical model to increase accuracy of de novo gene annotation, uniformity of read coverage, consistency of nucleotide frequencies and agreement with qRT-PCR | R | Anaconda, Bioconductor | ++ | http://master.bioconductor.org/packages/devel/bioc/html/seqbias.html | 1 |
| SAMStat[15] | 2011 | Post QC tool which plotsPost nucleotide overrepresentation and other statistics in mapped and unmapped reads in a html page | C | N/A | +++ | http://samstat.sourceforge.net | 1 |
| Samtools[16] | 2009 | Post QC tool using generic alignment format for storing read alignments against reference sequences and to visualize the Binary/Alignment Map (BAM). | C, Perl | Anaconda | + | https://github.com/samtools/samtools | 1 |
| **b. Read alignment** | | | | | | | |
| deSALT[17] | 2019 | Long transcriptomic read alignment with de Bruijn graph-based index | C | Anaconda | ++ | https://github.com/ydLiu-HIT/deSALT | 1 |
| Magic-BLAST[18] | 2018 | Aligner for long and short reads through optimization of a spliced alignment score | C++ | N/A | +++ | https://ncbi.github.io/magicblast/ | 1 |
| Minimap2[19] | 2018 | Alignment using seed chain alignment procedure | C, Python | Anaconda | ++ | https://github.com/lh3/minimap2 | 1 |
| DART[20] | 2018 | Burrows-Wheeler Transform based aligner which adopts partitioning strategy to divide a read into two groups | C/C++ | Anaconda | ++ | https://github.com/hsinnan75/DART | 1 |
| MMR[21] | 2016 | Resolves the mapping location of multi-mapping reads, optimising for locally smooth coverage. | C++ | N/A | +++ | https://github.com/ratschlab/mmr | 1 |
| ContextMap 2[22] | 2015 | Allows parallel mapping against several reference genomes | Java | N/A | +++ | http://www.bio.ifi.lmu.de/ContextMap | 2 |
| HISAT[23] | 2015 | Aligning reads using an indexing scheme based on the Burrows-Wheeler transform and the Ferragina-Manzini (FM) index | C++ | Anaconda | ++ | http://www.ccb.jhu.edu/software/hisat/index.shtml | 2 |
| Segemehl[24] | 2014 | Multi-split mapping for circular RNA, trans-splicing, and fusion events in addition to performing splice alignment | C, C++, Perl, Python, Shell (Bash) | Anaconda | ++ | (http://www.bioinf.uni-leipzig.de/Software/segemehl/). | 2 |
| JAGuaR[25] | 2014 | Uses a modified GTF (Gene Transfer Format) of known splice sites to build the complete sequence from all reads mapped to the transcript. | Python | N/A | +++ | https://www.bcgsc.ca/resources/software/jaguar | 2 |
| CRAC[26] | 2013 | Uses double K-mer indexing and profiling approach to map reads, predict SNPs, gene fusions, repeat borders. | C++ | Anaconda | ++ | http://crac.gforge.inria.fr/ | 2 |
| STAR[27] | 2013 | Aligns long reads against genome reference database | C++ | Anaconda | ++ | https://github.com/alexdobin/STAR | 1 |
| Subread[28] | 2013 | Mapping reads to a reference genome using multi-seed strategy, called seed-and-vote | C, R | Anaconda, Bioconductor | ++ | https://bioconductor.org/packages/release/bioc/html/Rsubread.html | 1 |
| TopHat2[29] | 2013 | Alignment of transcriptomes in the presence of insertions, deletions and gene fusions | C++, Python | Anaconda | ++ | http://ccb.jhu.edu/software/tophat/index.shtml | 2 |
| OSA[30] | 2012 | K-mer profiling approach to map reads | C# | N/A | +++ | http://www.arrayserver.com/wiki/index.php?title=OSA | 2 |
| PASSion[31] | 2012 | Pattern growth pipeline for splice junction detection | C++, Perl, Shell (Bash) | N/A | +++ | https://trac.nbic.nl/passion/ | 2 |
| RUM[32] | 2011 | Comparative analysis of RNA-seq alignment algorithms and the RNA-seq unified mapper | Perl, Python | N/A | +++ | http://www.cbil.upenn.edu/RUM/ | 2 |
| SOAPSplice[33] | 2011 | Ab initio detection of splice junctions | Perl | Anaconda | ++ | http://soap.genomics.org.cn/soapsplice.html | 2 |
| MapSplice[34] | 2010 | De novo detection of splice junctions | C++ | Anaconda | ++ | https://github.com/LiuBioinfo/MapSplice | 1 |
| SpliceMap[35] | 2010 | De novo detection of splice junctions and RNA-seq alignment | C++ | Anaconda | ++ | http://web.stanford.edu/group/wonglab/SpliceMap/ | 2 |
| Supersplat[36] | 2010 | De novo detection of splice junctions | C++ | N/A | +++ | http://mocklerlab.org/tools/1/manual | 2 |
| HMMSplicer[37] | 2010 | Detection of splice junctions of short sequence reads | Python | N/A | +++ | http://derisilab.ucsf.edu/software/hmmsplicer | 2 |
| QPALMA[38] | 2008 | Spliced alignments of short sequence reads. | C++, Python | N/A | +++ | http://www.raetschlab.org/suppl/qpalma | 2 |
| **c. Gene annotations** | | | | | | | |
| SQANTI[39] | 2018 | Analyses quality of long reads transcriptomes and removes artefacts. | Python | Anaconda | ++ | https://github.com/ConesaLab/SQANTI | 1 |
| Annocript[40] | 2015 | Databases are downloaded to annotate protein coding transcripts with the prediction of putative long non-coding RNAs in whole transcriptomes. | Perl, Python, R | N/A | +++ | https://github.com/frankMusacchia/Annocript | 1 |
| CIRI[41] | 2015 | De novo circular RNA identification | Perl | N/A | +++ | https://sourceforge.net/projects/ciri/ | 1 |
| TSSAR[42] | 2014 | Automated de novo TSS annotation from differential RNA-seq data | Java, Perl, R | Anaconda | ++ | http://rna.tbi.univie.ac.at/TSSAR | 2 |
| **d. Transcriptome assembly** | | | | | | | |
| FLAIR[43] | 2020 | Full-length alternative isoform analysis of RNA | Python | Anaconda | ++ | https://github.com/BrooksLabUCSC/FLAIR | 1 |
| Scallop[44] | 2017 | Splice-graph-decomposition algorithm which optimizes two competing objectives while satisfying all phasing constraints posed by reads spanning multiple vertices | C++ | Anaconda | +++ | https://github.com/Kingsford-Group/scallop | 1 |
| CLASS2[45] | 2016 | Splice variant annotation | C++, Perl, Shell | Anaconda | ++ | https://sourceforge.net/projects/splicebox/ | 1 |
| StringTie[46] | 2015 | Applies a network flow algorithm originally developed in optimization theory, together with optional de novo assembly, to assemble transcripts | C++ | N/A | +++ | http://ccb.jhu.edu/software/stringtie | 2 |
| Bridger[47] | 2015 | De novo transcript assembler using a mathematical | C++, Perl | N/A | +++ | https://sourceforge.net/projects/rnaseqassembly/files/?source=navbar | 1 |
| | | model, called the minimum path cover | | | | | |
| Bayesembler[48] | 2014 | Reference genome guided transcriptome assembly built on a Bayesian model | C++ | N/A | +++ | https://github.com/bioinformatics-centre/bayesembler. | 1 |
| SEECER[49] | 2013 | De novo transcriptome assembly using hidden Markov Model (HMM) based method | C++ | N/A | +++ | http://sb.cs.cmu.edu/seecer/ | 2 |
| BRANCH[50] | 2013 | De novo transcriptome assemblies by using genomic information that can be partial or complete genome sequences from the same or a related organism. | C++ | N/A | +++ | https://github.com/baoe/BRANCH | 1 |
| EBARDenovo[51] | 2013 | De novo transcriptome assembly uses an | C# | N/A | +++ | https://sourceforge.net/projects/ebardenovo/ | 1 |
| | | efficient chimera-detection function | | | | | |
| Oases[52] | 2012 | De novo transcriptome assembly using k-mer profiling and building a de Brujin graph | C | N/A | +++ | https://github.com/dzerbino/oases/tree/master | 1 |
| Cufflinks[53] | 2012 | Ab initio transcript assembly, estimates their abundances, and tests for differential expression | C++ | N/A | +++ | https://github.com/cole-trapnell-lab/cufflinks | 1 |
| IsoInfer[54] | 2011 | Infer isoforms from short reads | C/C++ | N/A | +++ | http://www.cs.ucr.edu/~jianxing/IsoInfer.html | 2 |
| IsoLasso[55] | 2011 | Reference genome guided using LASSO regression approach | C++ | N/A | +++ | http://alumni.cs.ucr.edu/~liw/isolasso.html | 2 |
| Trinity[56] | 2011 | De novo transcriptome assembly | C++, Java, Perl, R, | Anaconda | ++ | https://github.com/trinityrnaseq/trinityrnaseq/wiki | 1 |
| | | | Shell (Bash) | | | | |
| Trans-ABySS[57] | 2010 | De novo short-read transcriptome assembly and can also be used for fusion detection | Python | N/A | +++ | https://github.com/bcgsc/transabyss | 1 |
| Scripture[58] | 2010 | Ab initio reconstruction of transcriptomes of pluripotent and lineage committed cells | Java | N/A | +++ | www.broadinstitute.org/software/Scripture/ | 2 |
| **e. Transcriptome quantification** | | | | | | | |
| TALON[59] | 2019 | Long-read transcriptome discovery and quantification | Python | N/A | +++ | https://github.com/dewyman/TALON | 1 |
| Salmon[60] | 2017 | Composed of: lightweight-mapping model, an online phase that estimates initial expression levels and model parameters, and an offline phase | C++ | Anaconda | ++ | https://github.com/COMBINE-lab/Salmon | 1 |
| | | that refines expression estimates models, and mesures sequence-specific , fragment GC, and positional biases | | | | | |
| Kallisto[61] | 2016 | K-mer based pseudoalignment for allligment free transcript and gene expression quantification | C, C++, Perl | Anaconda | ++ | https://github.com/pachterlab/kallisto | 1 |
| Wub[62] | 2016 | Sequence and error simulation tool to calculate read and genome assembly accuracy. | Python | Anaconda | ++ | https://github.com/nanoporetech/wub | 1 |
| Rcount[63] | 2015 | GUI based tool used for quantification using counts per feature | Web based tool | N/A | + | https://github.com/MWSchmid/Rcount | 1 |
| Ht-seq[64] | 2015 | Calculates gene counts by counting number of reads overlapping genes | Python | pip | ++ | https://htseq.readthedocs.io/en/release_0.11.1/overview.html | 2 |
| EMSAR[65] | 2015 | Estimation by mappability-based segmentation and reclustering using a joint Poisson model | C | N/A | +++ | https://github.com/parklab/emsar | 1 |
| Maxcounts[66] | 2014 | Quantify the expression assigned to an exon as the maximum of its per-base counts | C++ | N/A | +++ | http://sysbiobig.dei.unipd.it/?q=Software#MAXCOUNTS | 2 |
| FIXSEQ[67] | 2014 | A nonparametric and universal method for processing per-base sequencing read count data. | R | N/A | ++ | https://bitbucket.org/thashim/fixseq/src/master/ | 1 |
| Sailfish[68] | 2014 | EM based quantification using statistical | C, C++ | Anaconda | ++ | https://github.com/kingsfordgroup/sailfish | 1 |
| | | coupling between k-mers. | | | | | |
| Casper[69] | 2014 | Bayesian modeling framework to quantify alternative splicing. | R | Anaconda, Bioconductor | ++ | http://www.bioconductor.org/packages/release/bioc/html/casper.html | 1 |
| MaLTA[70] | 2014 | Simultaneous transcriptome assembly and quantification from Ion Torrent RNA-Seq data. | C++ | N/A | +++ | http://alan.cs.gsu.edu/NGS/?q=malta | 2 |
| Featurecounts[71] | 2014 | Read summarization program for counting reads generated. | R | N/A | ++ | http://subread.sourceforge.net/ | 1 |
| MITIE[72] | 2013 | Transcript reconstruction and assembly from RNA-Seq data using mixed integer optimisation. | MATLAB, C++ | N/A | +++ | https://github.com/ratschlab/MiTie | 1 |
| iReckon[73] | 2013 | EM-based method to | Java | N/A | +++ | http://compbio.cs.toronto.edu/ireckon/ | 2 |
| | | accurately estimate the abundances of known and novel isoforms. | | | | | |
| eXpress[74] | 2013 | Online EM based algorithm for quantification which considers one read at a time. | C++, Shell (Bash) | Anaconda | ++ | https://pachterlab.github.io/eXpress/manual.html | 1 |
| BitSeq[75] | 2012 | Bayesian transcript expression quantification and differential expression. | C++, R | Anaconda,Bioconductor | ++ | http://bitseq.github.io/ | 1 |
| IQSeq[76] | 2012 | Integrated isoform quantification analysis. | C++ | N/A | +++ | http://archive.gersteinlab.org/proj/rnaseq/IQSeq/ | 2 |
| CEM[77] | 2012 | Statistical framework for both transcriptome assembly and | Python | N/A | +++ | http://alumni.cs.ucr.edu/~liw/cem.html | 2 |
| | | isoform expression level estimation. | | | | | |
| SAMMate[78] | 2011 | Analysis of differential gene and isoform expression. | Java | N/A | +++ | http://sammate.sourceforge.net/ | 1 |
| Isoformex[79] | 2011 | Estimation method to estimate the expression levels of transcript isoforms. | N/A | N/A | N/A | http://bioinformatics.wistar.upenn.edu/isoformex | 2 |
| IsoEM[80] | 2011 | EM based method for inference of isoform and gene-specific expression levels | Java | N/A | +++ | http://dna.engr.uconn.edu/software/IsoEM/. | 2 |
| RSEM[81] | 2011 | Ab initio EM based method for inference of isoform and gene-specific expression levels | C++, Perl, Python, R | Anaconda | ++ | https://github.com/deweylab/RSEM | 1 |
| EDASeq[82] | 2011 | Within-lane GC-content normalization, between-sample normalization, visualization. | R | Anaconda, Bioconductor | ++ | https://bioconductor.org/packages/devel/bioc/html/EDASeq.html | 1 |
| MMSEQ[83] | 2011 | Haplotype and isoform specific expression estimation | C++, R, Ruby, Shell (Bash) | N/A | ++ | https://github.com/eturro/mmseq | 1 |
| MISO[84] | 2010 | Statistical model that estimates expression of alternatively spliced exons and isoforms | C, Python | N/A | +++ | https://miso.readthedocs.io/en/fastmiso/#latest-version-from-github | 2 |
| SOLAS[85] | 2010 | Prediction of alternative isoforms from exon expression levels | R | N/A | ++ | http://cmb.molgen.mpg.de/2ndGenerationSequencing/Solas/ | 2 |
| Rseq[86] | 2009 | Statistical inferences for isoform expression | C++ | N/A | +++ | http://www-personal.umich.edu/~jianghui/rseq/#download | 2 |
| rQuant[87] | 2009 | Estimating density biases and considering the read coverages at each nucleotide independently using quadratic programming | Matlab, Shell (Bash), Javascript | N/A | +++ | https://galaxy.inf.ethz.ch/?tool_id=rquantweb&version=2.2&__identifer=3iuqb8nb3wf | 2 |
| ERANGE[88] | 2008 | Mapping and quantifying mammalian transcripts | Python | N/A | +++ | http://woldlab.caltech.edu/rnaseq | 2 |
| **f. Differential expression** | | | | | | | |
| Swish[89] | 2019 | Non-parametric model for differential expression analysis using inferential replicate counts | R | Bioconductor | ++ | https://bioconductor.org/packages/release/bioc/html/fishpond.html | 1 |
| Yanagi[90] | 2019 | Transcriptome segment analysis | Python/C++ | N/A | +++ | https://github.com/HCBravoLab/yanagi | 1 |
| Whippet[91] | 2018 | Quantification of transcriptome | Julia | N/A | +++ | https://github.com/timbitz/Whippet.jl | 1 |
| | | structure and gene expression analysis using EM. | | | | | |
| ReQTL[92] | 2018 | Identifies correlations between SNVs and gene expression from RNA-seq data | R | N/A | ++ | https://github.com/HorvathLab/ReQTL | 1 |
| vast-tools[93] | 2017 | Profiling and comparing alternative splicing events in RNA-Seq data and for downstream analyses of alternative splicing. | R, Perl | N/A | +++ | https://github.com/vastgroup/vast-tools | 1 |
| Ballgown[94] | 2015 | Linear model–based differential expression analyses | R | Anaconda, Bioconductor | ++ | https://github.com/alyssafrazee/ballgown | 1 |
| Limma/Voom[95] | 2014 | Linear model-based differential expression and differential splicing analyses | R | Anaconda, Bioconductor | ++ | https://bioconductor.org/packages/release/bioc/html/limma.html | 1 |
| rMATS[96] | 2014 | Detect major differential alternative splicing types in RNA-seq data with replicates. | Python, C++ | Anaconda | ++ | http://rnaseq-mats.sourceforge.net/rmats3.2.5/ | 1 |
| DESeq2[97] | 2014 | Differential analysis of count data, using shrinkage estimation for dispersions and fold changes | R | Bioconductor, CRAN | ++ | https://bioconductor.org/packages/release/bioc/html/DESeq2.html | 1 |
| Corset[98] | 2014 | Differential gene expression analysis for de novo assembled transcriptomes | C++ | Anaconda | ++ | https://github.com/Oshlack/Corset/wiki | 1 |
| BADGE[99] | 2014 | Bayesian model for accurate abundance | Matlab | N/A | +++ | http://www.cbil.ece.vt.edu/software.htm | 2 |
| | | quantification and differential analysis | | | | | |
| compcodeR[100] | 2014 | Benchmarking of differential expression analysis methods | R | Anaconda, Bioconductor | ++ | https://www.bioconductor.org/packages/compcodeR/ | 1 |
| metaRNASeq[101] | 2014 | Differential meta-analyses of RNA-seq data | R | Anaconda, CRAN | ++ | http://cran.r-project.org/web/packages/metaRNASeq | 1 |
| Characteristic Direction[102] | 2014 | Geometrical multivariate approach to identify differentially expressed genes | R, Python, MATLAB | N/A | ++ | http://www.maayanlab.net/CD | 2 |
| HTSFilter[103] | 2013 | Filter-replicated high-throughput transcriptome sequencing data | R | Anaconda, Bioconductor | ++ | http://www.bioconductor.org/packages/release/bioc/html/HTSFilter.html | 1 |
| NPEBSeq[104] | 2013 | Nonparametric empirical bayesian-based procedure for differential expression analysis | R | N/A | ++ | http://bioinformatics.wistar.upenn.edu/NPEBseq | 2 |
| EBSeq[105] | 2013 | Identifying differentially expressed isoforms. | R | Anaconda, Bioconductor | ++ | http://bioconductor.org/packages/release/bioc/html/EBSeq.html | 1 |
| sSeq[106] | 2013 | Shrinkage estimation of dispersion in Negative Binomial models | R | Anaconda, Bioconductor | ++ | http://bioconductor.org/packages/release/bioc/html/sSeq.html | 1 |
| Cuffdiff2[107] | 2013 | Differential analysis at transcript resolution | C++, Python | N/A | +++ | http://cole-trapnell-lab.github.io/cufflinks/cuffdiff/ | 1 |
| SAMseq[108] | 2013 | Nonparametric method with resampling to account for the different sequencing depths | R | CRAN | ++ | https://rdrr.io/cran/samr/man/SAMseq.html | 1 |
| DSGseq [109] | 2013 | NB-statistic method that can detect differentially spliced genes between two groups of | R | N/A | ++ | http://bioinfo.au.tsinghua.edu.cn/software/DSGseq/ | 2 |
| | | samples without using a prior knowledge on the annotation of alternative splicing. | | | | | |
| NOISeq[110] | 2011 | Uses a non-parametric approach for differential expression analysis and can work in absence of replicates | R | Bioconductor | ++ | https://bioconductor.org/packages/release/bioc/html/NOISeq.html | 1 |
| EdgeR[111] | 2010 | Examining differential expression of replicated count data and differential exon usage | R | Anaconda, Bioconductor | ++ | https://bioconductor.org/packages/release/bioc/html/edgeR.html | 1 |
| DEGseq[112] | 2010 | Identify differentially expressed genes or isoforms for RNA-seq data from different samples. | R | Bioconductor | ++ | http://bioconductor.org/packages/release/bioc/html/DEGseq.html | 1 |

| g. RNA splicing | | | | | | | |
|---|---|---|---|---|---|---|---|
| LeafCutter[113] | 2018 | Detects differential splicing and maps quantitative trait loci (sQTLs). | R | Anaconda | ++ | https://github.com/davidaknowles/leafcutter.git | 1 |
| MAJIQ-SPEL[114] | 2018 | Visualization, interpretation, and experimental validation of both classical and complex splicing variation and automated RT-PCR primer design. | C++, Python | N/A | +++ | **https://galaxy.biociphers.org/galaxy/root?tool_id=majiq_spel** | 2 |
| MAJIQ[115] | 2016 | Web-tool that takes as input local splicing variations (LSVs) quantified from RNA-seq data and provides users with a visualization package (VOILA) and quantification of gene isoforms. | C++, Python | N/A | +++ | https://majiq.biociphers.org/commercial.php | 2 |
| SplAdder[116] | 2016 | Identification, quantification, and testing of alternative splicing events | Python | N/A | +++ | http://github.com/ratschlab/spladder | 1 |
| SplicePie[117] | 2015 | Detection of alternative, non-sequential and recursive splicing | Perl, R | N/A | +++ | https://github.com/pulyakhina/splicing_analysis_pipeline | 1 |
| SUPPA[118] | 2015 | Alternative splicing analysis | Python, R | Anaconda | ++ | https://github.com/comprna/SUPPA | 1 |
| SNPlice[119] | 2015 | Identifying variants that modulate Intron retention | Python | N/A | +++ | https://code.google.com/p/snplice/ | 1 |
| IUTA[120] | 2014 | Detecting differential isoform usage | R | N/A | ++ | http://www.niehs.nih.gov/research/resources/software/biostatistics/iuta/index.cfm. | 1 |
| SigFuge[121] | 2014 | Identifying genomic loci exhibiting differential | R | Anaconda, Bioconductor | ++ | http://bioconductor.org/packages/release/bioc/html/SigFuge.html | 1 |
| | | transcription patterns | | | | | |
| FineSplice[122] | 2014 | Splice junction detection and quantification | Python | N/A | +++ | https://sourceforge.net/p/finesplice/ | 1 |
| PennSeq[123] | 2014 | Statistical method that allows each isoform to have its own non-uniform read distribution | Perl | N/A | +++ | http://sourceforge.net/projects/pennseq | 1 |
| FlipFlop[124] | 2014 | RNA isoform identification and quantification with network flows | R | Anaconda, Bioconductor | ++ | https://bioconductor.org/packages/release/bioc/html/flipflop.html | 1 |
| GESS[125] | 2014 | Graph-based exon-skipping scanner for de novo detection of skipping event sites | N/A | N/A | N/A | http://jinlab.net/GESS_Web/ | 2 |
| spliceR[126] | 2013 | Classification of alternative splicing and prediction of coding potential | R | Anaconda, Bioconductor | ++ | http://www.bioconductor.org/packages/2.13/bioc/html/spliceR.html | 1 |
| RNASeq-MATS[127] | 2013 | Detects and analyzes differential alternative splicing events | C, Python | N/A | +++ | http://rnaseq-mats.sourceforge.net/ | 1 |
| SplicingCompass[128] | 2013 | Differential splicing detection | R | N/A | ++ | http://www.ichip.de/softwa | 2 |
| DiffSplice[129] | 2013 | Genome-wide detection of differential splicing | C++ | N/A | +++ | http://www.netlab.uky.edu/p/bioinfo/DiffSplice | 2 |
| DEXSeq[130] | 2012 | Statistical method to test for differential exon usage. | R | Anaconda, Bioconductor | ++ | https://bioconductor.org/packages/release/bioc/html/DEXSeq.html | 1 |
| SpliceSeq[131] | 2012 | Identifies differential splicing events between test and control groups. | Java | N/A | +++ | http://bioinformatics.mdanderson.org/main/SpliceSeq:Overview. | 2 |
| JuncBASE[132] | 2011 | Identification and quantification of alternative splicing, including | Python | N/A | +++ | https://github.com/anbrooks/juncBASE | 1 |
| | | unannotated splicing | | | | | |
| ALEXA-seq[133] | 2010 | Alternative expression analysis. | Perl, R, Shell (Bash) | N/A | +++ | http://www.alexaplatform.org/alexa_seq/. | 1 |
| **h. Cell deconvolution** | | | | | | | |
| TIMER2.0[134] | 2020 | Web server for comprehensive analysis of Tumor-Infiltrating Immune Cells. | Web-tool R, Javascript | N/A | + | https://github.com/taiwenli/TIMER | 1 |
| CIBERSORTx[135] | 2019 | Impute gene expression profiles and provide an estimation of the abundances of member cell types in a mixed cell population. | Web-tool Java, R | N/A | + | https://cibersortx.stanford.edu/ | 2 |
| quanTIseq[136] | 2019 | Quantify the fractions of ten immune cell types from bulk RNA-sequencing data. | R, Shell (Bash) | N/A | + | https://icbi.i-med.ac.at/quantiseq | 2 |
| Immunedeconv[137] | 2019 | Benchmarking of transcriptome-based cell-type quantification methods for immuno-oncology | R | Anaconda | ++ | https://github.com/icbi-lab/immunedeconv | 1 |
| Linseed[138] | 2019 | Deconvolution of cellular mixtures based on linearity of transcriptional signatures. | C++, R | N/A | ++ | https://github.com/ctlab/LinSeed | 1 |
| deconvSEQ[139] | 2019 | Deconvolution of cell mixture distribution based on a generalized linear model. | R | N/A | ++ | https://github.com/rosedu1/deconvSeq | 1 |
| CDSeq[140] | 2019 | Simultaneously estimate both cell-type proportions and cell-type-specific expression profiles. | MATLAB, R | N/A | ++ | https://github.com/kkang7/CDSeq_R_Package | 1 |
| Dtangle[141] | 2019 | Estimates cell type proportions using publicly | R | Anaconda,CRAN | ++ | https://github.com/gjhunt/dtangle | 1 |
| | | available, often cross-platform, reference data. | | | | | |
| GEDIT[142] | 2019 | Estimate cell type abundances. | Web based tool Python, R | N/A | + | http://webtools.mcdb.ucla.edu/ | 2 |
| SaVant[143] | 2017 | Web based tool for sample level visualization of molecular signatures in gene expression profiles. | Javascript, R | N/A | +++ | http://newpathways.mcdb.ucla.edu/savant | 2 |
| EPIC[144] | 2017 | Simultaneously estimates the fraction of cancer and immune cell types. | R | N/A | ++ | http://epic.gfellerlab.org/ | 2 |
| WSCUnmix[145] | 2017 | Automated deconvolution of structured mixtures. | MATLAB | N/A | +++ | https://github.com/tedroman/WSCUnmix | 1 |
| Infino[146] | 2017 | Deconvolves bulk RNA-seq into cell type abundances and captures gene expression variability in a Bayesian model to measure deconvolution uncertainty. | R, Python | N/A | ++ | https://github.com/hammerlab/infino | 1 |
| MCP-counter[147] | 2016 | Estimating the population abundance of tissue-infiltrating immune and stromal cell populations. | R | N/A | + | https://github.com/ebecht/MCPcounter | 1 |
| CellCode[148] | 2015 | Latent variable approach to differential expression analysis for heterogeneous cell populations. | R | N/A | ++ | http://www.pitt.edu/~mchikina/CellCODE/ | 2 |
| PERT[149] | 2012 | Probabilistic expression | MATLAB | N/A | +++ | https://github.com/gquon/PERT | 1 |
| | | deconvolution method. | | | | | |
| **i. Immune repertoire profiling** | | | | | | | |
| ImReP[150] | 2018 | Profiling immunoglobulin repertoires across multiple human tissues. | Python | N/A | +++ | https://github.com/mandricigor/imrep/wiki | 1 |
| TRUST (T cell)[151] | 2016 | Landscape of tumor-infiltrating T cell repertoire of human cancers. | Perl | N/A | +++ | https://github.com/liulab-dfci/TRUST4 | 1 |
| V'DJer [152] | 2016 | Assembly-based inference of B-cell receptor repertoires from short reads with V'DJer. | C, C++ | Anaconda | ++ | https://github.com/mozack/vdjer | 1 |
| IgBlast-based pipeline[153] | 2016 | Statistical inference of a convergent antibody repertoire response. | C++ | N/A | +++ | https://www.ncbi.nlm.nih.gov/igblast/ | 1 |
| MiXCR[154] | 2015 | Processes big immunome data from raw sequences to quantitated clonotypes. | Java | Anaconda | ++ | https://github.com/milaboratory/mixcr | 1 |
| **j. Allele specific expression** | | | | | | | |
| EAGLE[155] | 2017 | Bayesian model for identifying GxE interactions based on associations between environmental variables and allele-specific expression. | C++, R | N/A | ++ | https://github.com/davidaknowles/eagle | 1 |
| ANEVA-DOT/ ANEVA[156] | 2019 | Identify ASE outlier genes / Quantify genetic variation in gene dosage from ASE data. | R | N/A | ++ | https://github.com/PejLab/ANEVA-DOT | 1 |
| aFC[157] | 2017 | Quantifying the regulatory effect | Python | N/A | +++ | https://github.com/secastel/aFC | 1 |
| | | size of cis-acting genetic variation | | | | | |
| phASER[158] | 2016 | Uses readback phasing to produce haplotype level ASE data (as opposed to SNP level) | Python | N/A | +++ | https://github.com/secastel/phaser | 1 |
| RASQUAL[159] | 2016 | Maps QTLs for sequenced based cellular traits by combining population and allele-specific signals. | C, R | N/A | +++ | https://github.com/natsuhiko/rasqual | 1 |
| allelecounter[160] | 2015 | Generate ASE data from RNAseq data and a genotype file. | Python | N/A | +++ | https://github.com/secastel/allelecounter | 1 |
| WASP[161] | 2015 | Unbiased allele-specific read mapping and discovery of molecular QTLs | C Python | Anaconda | ++ | https://github.com/bmvdgeijn/WASP/ | 1 |
| Mamba[162] | 2015 | Compares different patterns | R | N/A | ++ | http://www.well.ox.ac.uk/~rivas/mamba/. | 2 |
| | | of ASE across tissues | | | | | |
| MBASED [163] | 2014 | Allele-specific expression detection in cancer tissues and cell lines | R | Anaconda, Bioconductor | ++ | https://bioconductor.org/packages/release/bioc/html/MBASED.html | 1 |
| Allim[164] | 2013 | Estimates allele-specific gene expression. | Python, R | N/A | +++ | https://sourceforge.net/projects/allim/ | 1 |
| AlleleSeq[165] | 2011 | Identifies allele-specific events in mapped reads between maternal and paternal alleles. | Python, Shell | N/A | +++ | http://alleleseq.gersteinlab.org/ | 2 |
| **k. Viral detection** | | | | | | | |
| ROP[166] | 2018 | Dumpster diving in RNA-sequencing to find the source of 1 trillion reads across diverse adult human tissues | Python, Shell (Bash) | Anaconda | ++ | https://github.com/smangul1/rop | 1 |
| RNA CoMPASS[167] | 2014 | Simultaneous analysis of transcriptomes and metatranscriptomes from diverse biological specimens. | Perl, Shell, Java | N/A | ++ | http://rnacompass.sourceforge.net/ | 1 |
| VirusSeq[168] | 2013 | Identify viruses and their integration sites using next-generation sequencing of human cancer tissues | Perl, Shell (Bash) | N/A | +++ | http://odin.mdacc.tmc.edu/~xsu1/VirusSeq.html | 2 |
| VirusFinder[169] | 2013 | Detection of Viruses and Their Integration Sites in Host Genomes through Next Generation Sequencing Data | Perl | N/A | +++ | http://bioinfo.mc.vanderbilt.edu/VirusFinder/ | 2 |
| **l. Fusion detection** | | | | | | | |
| INTEGRATE-Vis[170] | 2017 | Generates plots focused on annotating each | Python | N/A | +++ | https://github.com/ChrisMaherLab/INTEGRATE-Vis | 1 |
| | | gene fusion at the transcript- and protein-level and assessing expression across samples. | | | | | |
| INTEGRATE-Neo[171] | 2017 | Gene fusion neoantigen discovery tool, which uses RNA-Seq reads and is capable of reporting tumor-specific peptides recognizable by immune molecules. | Python, C++ | N/A | +++ | https://github.com/ChrisMaherLab/INTEGRATE-Neo | 1 |
| INTEGRATE[172] | 2016 | Capable of integrating aligned RNA-seq and WGS reads and characterizes the quality of predictions. | C++ | N/A | +++ | https://sourceforge.net/projects/integrate-fusion/ | 1 |
| TRUP[173] | 2015 | Combines split-read and read-pair analysis | C++, Perl, R | N/A | +++ | https://github.com/ruping/TRUP | 1 |
| | | with de novo regional assembly for the identification of chimeric transcripts in cancer specimens. | | | | | |
| PRADA[174] | 2014 | Detect gene fusions but also performs alignments, transcriptome quantification; mainly integrated genome/transcriptome read mapping. | Python | Anaconda | ++ | http://sourceforge.net/projects/prada/ | 1 |
| Pegasus[175] | 2014 | Annotation and prediction of biologically functional gene fusion candidates. | Java, Perl, Python, Shell (Bash) | N/A | +++ | https://github.com/RabadanLab/Pegasus | 1 |
| FusionCatcher[176] | 2014 | Finding somatic fusion genes | Python | Anaconda | ++ | https://sourceforge.net/projects/fusioncatcher/ | 1 |
| FusionQ[177] | 2013 | Gene fusion detection and quantification from paired-end RNA-seq | C++, Perl, R | N/A | +++ | http://www.wakehealth.edu/CTSB/Software/Software.htm | 2 |
| Barnacle[178] | 2013 | Detecting and characterizing tandem duplications and fusions in de novo transcriptome assemblies | Python, Perl | N/A | +++ | http://www.bcgsc.ca/platform/bioinfo/software/barnacle | 2 |
| Dissect[179] | 2012 | Detection and characterization of structural alterations in transcribed sequences | C | N/A | +++ | http://dissect-trans.sourceforge.net | 1 |
| BreakFusion[180] | 2012 | Targeted assembly-based identification of gene fusions | C++, Perl | N/A | +++ | https://bioinformatics.mdanderson.org/public-software/breakfusion/ | 2 |
| EricScript[181] | 2012 | Identification of gene fusion products in | Perl, R, Shell (Bash) | Anaconda | ++ | http://ericscript.sourceforge.net | 1 |
| | | paired-end RNA-seq data. | | | | | |
| Bellerophontes[182] | 2012 | Chimeric transcripts discovery based on fusion model. | Java, Perl, Shell (Bash) | N/A | +++ | http://eda.polito.it/bellerophontes/ | 2 |
| GFML[183] | 2012 | Standard format for organizing and representing the significant features of gene fusion data. | XML | N/A | + | http://code.google.com/p/gfml-prototype/ | 1 |
| FusionHunter[184] | 2011 | Identifies fusion transcripts from transcriptional analysis. | C++ | N/A | +++ | https://github.com/ma-compbio/FusionHunter | 1 |
| ChimeraScan[185] | 2011 | Identifying chimeric transcription. | Python | Anaconda | ++ | https://code.google.com/archive/p/chimerascan/downloads | 1 |
| TopHat-fusion[186] | 2011 | Discovery of novel fusion transcripts. | C++, Python | N/A | +++ | http://ccb.jhu.edu/software/tophat/fusion_index.shtml | 2 |
| deFuse[187] | 2011 | Fusion discovery in tumor RNA-seq data. | C++, Perl, R | Anaconda | ++ | https://github.com/amcpherson/defuse/blob/master/README.md | 1 |
| **m. Detecting circRNA** | | | | | | | |
| CIRIquant[188] | 2020 | Accurate quantification and differential expression analysis of circRNAs. | Python | N/A | ++ | https://sourceforge.net/projects/ciri/files/CIRIquant | 1 |
| CIRI-vis[189] | 2020 | Visualization of circRNA structures. | Java | N/A | +++ | https://sourceforge.net/projects/ciri/files/CIRI-vis | 1 |
| Ularcirc[190] | 2019 | Analysis and visualisation of canonical and back splice junctions. | R | Bioconductor | ++ | https://github.com/VCCRI/Ularcirc | 1 |
| CLEAR[191] | 2019 | Circular and Linear RNA expression analysis. | Python | N/A | +++ | https://github.com/YangLab/CLEAR | 1 |
| CIRI-full[192] | 2019 | Reconstruct and quantify full-length circular RNAs. | Java | N/A | +++ | https://sourceforge.net/projects/ciri/files/CIRI-full | 1 |
| circAST[193] | 2019 | Full-length assembly and quantification of alternatively | Python | N/A | +++ | https://github.com/xiaofengsong/CircAST | 1 |
| | | spliced isoforms in Circular RNAs | | | | | |
| CIRI2[194] | 2018 | Denovo circRNA identification | Pearl | N/A | +++ | https://sourceforge.net/projects/ciri/files/CIRI2 | 1 |
| Sailfish-cir[195] | 2017 | Quantification of circRNAs using model-based framework | Python | N/A | +++ | https://github.com/zerodel/sailfish-cir | 1 |
| CircComPara[196] | 2017 | Multi-method detection of circRNAs | R, Python | N/A | +++ | http://github.com/egaffo/CirComPara | 1 |
| UROBORUS[197] | 2016 | Computationally identifying circRNAs from RNA-seq data | Perl | N/A | +++ | https://github.com/WGLab/UROBORUS/tree/master/bin | 1 |
| PTESFinder[198] | 2016 | Identification of non-co-linear transcripts | Shell, Java | N/A | +++ | https://sourceforge.net/projects/ptesfinder-v1/ | 1 |
| NCLscan[199] | 2016 | identification of non-co-linear transcripts (fusion, trans-splicing and circular RNA) | Python | N/A | +++ | https://github.com/TreesLab/NCLscan | 1 |
| DCC[200] | 2016 | Specific identification and | Python | N/A | +++ | https://github.com/dieterich-lab/DCC | 1 |
| | | quantification of circRNA | | | | | |
| CIRI-AS[201] | 2016 | Identification of internal structure and alternative splicing events in circRNA | Perl | N/A | +++ | https://sourceforge.net/projects/ciri/files/CIRI-AS | 1 |
| circTest[200] | 2016 | Differential expression analysis and plotting of circRNAs | R | N/A | +++ | https://github.com/dieterich-lab/CircTest | 1 |
| CIRCexplorer2[202] | 2016 | Annotation and de novo assembly of circRNAs | Python | Anaconda | ++ | https://github.com/YangLab/CIRCexplorer2 | 1 |
| KNIFE [203] | 2015 | Statistically based detection of circular and linear isoforms from RNA-seq data | Perl, Python, R | N/A | +++ | https://github.com/lindaszabo/KNIFE | 1 |
| circRNA_finder[204] | 2014 | Identification of circRNAs from RNA-seq data | Perl | N/A | +++ | https://github.com/orzechoj/circRNA_finder | 1 |
| find_circ[205] | 2013 | Identification of circRNAs based | Python | N/A | ++ | https://github.com/marvin-jens/find_circ | 1 |
| | | on head-to-tail spliced sequencing reads | | | | | |
| **n. Small RNA detection** | | | | | | | |
| miRTrace[206] | 2018 | Quality control of miRNA-seq data, identifies cross-species contamination. | Java | Anaconda | +++ | https://github.com/friedlanderlab/mirtrace | 1 |
| sRNAbench[207] | 2015 | Expression profiling of small RNAs, prediction of novel microRNAs, analysis of isomiRs, genome mapping and read length statistics. | Web based tool | N/A | + | https://bioinfo5.ugr.es/srnatoolbox/srnabench/ | 2 |
| sRNAde[207] | 2015 | Detection of differentially expressed small RNAs based on three programs. | Web based tool | N/A | + | https://bioinfo5.ugr.es/srnatoolbox/srnade/ | 2 |
| sRNAblast[207] | 2015 | Aimed to determine the origin of unmapped or | Web based tool | N/A | + | https://bioinfo5.ugr.es/srnatoolbox/srnablast/ | 2 |
| | | unassigned reads by means of a blast search against several remote databases. | | | | | |
| miRNAconsTarget[207] | 2015 | Consensus target prediction on user provided input data. | Web based tool | N/A | + | https://bioinfo5.ugr.es/srnatoolbox/amirconstarget/ | 2 |
| sRNAjBrowser[207] | 2015 | Visualization of sRNA expression data in a genome context. | Web based tool | N/A | + | https://bioinfo5.ugr.es/srnatoolbox/srnajbrowser/ | 2 |
| sRNAjBrowserDE[207] | 2015 | Visualization of differential expression as a function of read length in a genome context. | Web based tool | N/A | + | https://bioinfo5.ugr.es/srnatoolbox/srnajbrowserde/ | 2 |
| ShortStack[208] | 2013 | Analyzes reference-aligned sRNA-seq data and performs. comprehensive *de novo* annotation and quantification of the inferred sRNA genes. | Perl | Anaconda | +++ | https://github.com/MikeAxtell/ShortStack | 1 |
| mirTools 2.0[209] | 2013 | Detect, identify and profile various types, functional annotation and differentially expressed sRNAs. | Web based tool | N/A | + | http://www.wzgenomics.cn/mr2_dev/ | 2 |
| UEA sRNA Workbench[210] | 2012 | Complete analysis of single or multiple-sample small RNA datasets. | Web based tool, C++, Java | N/A | +++ | https://sourceforge.net/projects/srnaworkbench/ | 1 |
| miRDeep2[211] | 2011 | Discovers known and novel miRNAs, quantifies miRNA expression. | Perl | Anaconda | +++ | https://github.com/rajewsky-lab/mirdeep2 | 1 |
| miRanalyzer[212] | 2011 | Detection of known and prediction of new microRNAs in high-throughput sequencing experiments. | Web based tool | N/A | + | http://bioinfo2.ugr.es/miRanalyzer/miRanalyzer.php | 2 |
| SeqBuster[213] | 2010 | Provides an automatized pre-analysis for sequence annotation for analysing small RNA data from Illumina sequencing. | Web based tool | Anaconda | + | http://estivill_lab.crg.es/seqbuster | 2 |
| DARIO[214] | 2010 | Allows to study short read data and provides a wide range of analysis features, including quality control, read normalization, and quantification. | Web based tool | N/A | + | http://dario.bioinf.uni-leipzig.de/index.py | 2 |
| **o. Visualization tools** | | | | | | | |
| BEAVR[215] | 2020 | Facilitates interactive analysis and exploration of RNA-seq data, allowing statistical testing and visualization of the table of | R | N/A | ++ | https://github.com/developerpiru/BEAVR | 1 |
| | | differentially expressed genes obtained. | | | | | |
| coseq[216] | 2018 | Co-expression analysis of sequencing data | R | Anaconda, Bioconductor | ++ | https://bioconductor.org/packages/release/bioc/html/coseq.html | 1 |
| ReadXplorer[217] | 2016 | Read mapping analysis and visualization | Java | N/A | +++ | https://www.uni-giessen.de/fbz/fb08/Inst/bioinformatik/software/ReadXplorer | 2 |
| Integrated Genome Browser[218] | 2016 | An interactive tool for visually analyzing tiling array data and enables quantification of alternative splicing | Java | N/A | +++ | http://www.bioviz.org/ | 2 |
| Sashimi plots[219] | 2015 | Quantitative visualization comparison of exon usage | Python | N/A | ++ | http://miso.readthedocs.org/en/fastmiso/sashimi.html | 2 |
| ASTALAVISTA[220] | 2015 | Reports all alternative splicing events reflected by transcript annotations | Java | Anaconda | ++ | http://astalavista.sammeth.net/ | 2 |
| RNASeqBrowser[221] | 2015 | Incorporates and extends the functionality of the UCSC genome browser | Java | N/A | +++ | http://www.australianprostatecentre.org/research/software/rnaseqbrowser | 2 |
| SplicePlot[222] | 2014 | Visualizing splicing quantitative trait loci | Python | N/A | +++ | http://montgomerylab.stanford.edu/spliceplot/index.html | 2 |
| RNASeqViewer[223] | 2014 | Compare gene expression and alternative splicing | Python | N/A | +++ | https://sourceforge.net/projects/rnaseqbrowser/ | 1 |
| PrimerSeq[224] | 2014 | Systematic design and visualization of RT-PCR primers using RNA seq data | Java, C++, Python | N/A | +++ | http://primerseq.sourceforge.net/ | 1 |
| Epiviz[225] | 2014 | Combining algorithmic-statistical analysis and interactive visualization | R | Anaconda, Bioconductor | ++ | https://epiviz.github.io/ | 1 |
| RNAbrowse[226] | 2014 | RNA-seq De Novo Assembly Results Browser | N/A | N/A | N/A | http://bioinfo.genotoul.fr/RNAbrowse | 2 |
| ZENBU[227] | 2014 | Interactive visualization and analysis of large-scale sequencing datasets | C++, Javascript | N/A | + | https://fantom.gsc.riken.jp/zenbu/ | 2 |
| CummeRbund[53] | 2012 | Navigate through data produced from a Cuffdiff RNA-seq differential expression analysis | R | Anaconda, Bioconductor | ++ | http://bioconductor.org/packages/devel/bioc/html/cummeRbund.html | 1 |
| Splicing Viewer[228] | 2012 | Visualization of splice junctions and alternative splicing | Java | N/A | +++ | http://bioinformatics.zj.cn/splicingviewer. | 2 |

**Table 1:** Landscape of current computational methods for RNA-seq analysis. We categorized RNA-seq tools published from 2008 to 2020 based on processes in the RNA-seq pipeline and workflow; starting with data quality control, read alignment, gene annotations, transcriptome assembly, transcriptome quantification, differential expression, RNA splicing, cell deconvolution, immune repertoire profiling, allele specific expression, viral detection, fusion detection, detecting circRNA, small RNA detection, and visualization tools. The third column ("Notable Features") presents key functionalities and methods used. The fourth column ("Programing Language")

presents the interface mode (e.g., GUI, web-based, programming language). The fifth column (“Package Manager”) highlights if a package manager such as Anaconda, Bioconductor, CRAN, Docker Hub, pip, or PyPI is available for the tool. We designated the assumed expertise level with a +, ++, or +++ in the sixth column (“Required Expertise”). A “+” represents little to no required expertise which would be assigned to a GUI based/web interface tool. “++” was assigned to tools that require R and/or multiple programming languages and whose software is located on Anaconda, Bioconductor, CRAN, Docker Hub, pip, or PyPI. “+++” was assigned to tools that require expertise in languages such as C, C++, Java, Python, Perl, or Shell (Bash) and may or may not have a package manager present. For each tool, we provide the links where the published tool software can be found and downloaded (“Software”). In the seventh column (“Type of URL”), each tool was assigned a “1” for web services designed to host source code or “2” for others (e.g personal and/or university web services).